\documentclass[aps,twocolumn,nofootinbib,groupedaddress,amsfonts,floatfix]{revtex4-1} 
\usepackage{graphicx,amsmath,amssymb,amstext}
\usepackage{amssymb,amsbsy,amsfonts,amsthm,color}

\usepackage{epsfig}
\usepackage{graphicx}
\usepackage{subfigure}
\usepackage{hyperref}
\usepackage{cancel}
\usepackage{balance}

\graphicspath{{Figures/}}

\usepackage{color}
\usepackage[dvipsnames]{xcolor}

\renewcommand{\eqref}[1]{Eqn. (\ref{#1})}
\newcommand{\Beq}{\begin{eqnarray}}
\newcommand{\Eeq}{\end{eqnarray}}

\def\lsim{\mathrel {\vcenter {\baselineskip 0pt \kern 0pt \hbox{$<$} \kern 0pt \hbox{$\sim$} }}}

\newcommand{\mpl}{M_{\mbox{\tiny Pl}}}

\def\gsim{\mathrel {\vcenter {\baselineskip 0pt \kern 0pt \hbox{$>$} \kern 0pt \hbox{$\sim$} }}}

\newcommand{\eqn}[1]{Eqn. \ref{#1}}

\newcommand{\threedsquaremat}[9]{\left( \begin{array}{ccc}#1 & #2 & #3\\ #4 & #5 & #6 \\ #7 & #8 & #9 \end{array} \right)}

\begin{document}
{\hfill KCL-PH-TH/2020-06}
\title{Coherent Gravitational Waveforms and Memory  from Cosmic String Loops }

\author{Josu C. Aurrekoetxea ${^a}$}
\email{j.c.aurrekoetxea@gmail.com}

\author{Thomas Helfer${^a}{^b}$}
\email{thomashelfer@live.de}

\author{Eugene A. Lim${^a}$}
\email{eugene.a.lim@gmail.com}

\affiliation{\vspace{0.2cm}${^a}$Theoretical Particle Physics and Cosmology Group, Physics Department,
Kings College London, Strand, London WC2R 2LS, United Kingdom}
\affiliation{\vspace{0.05cm}${^b}$Department of Physics and Astronomy, Johns Hopkins University, Baltimore, MD 21218, USA}

\begin{abstract}
We construct, for the first time, the time-domain gravitational wave strain waveform from the collapse of a strongly gravitating Abelian Higgs cosmic string loop in full general relativity.  We show that the strain exhibits a large memory effect during merger, ending with a burst and the characteristic ringdown as a black hole is formed. Furthermore, we investigate the waveform and energy  emitted as a function of string width, loop radius and  string tension $G\mu$. We  find that the mass normalized gravitational wave energy displays a strong dependence on the inverse of the string tension $E_{\mathrm{GW}}/M_0\propto 1/G\mu$, with $E_{\mathrm{GW}}/M_0 \sim {\cal O}(1)\%$ at the percent level, for the regime where $G\mu\gtrsim10^{-3}$. Conversely, we show that the efficiency is only weakly dependent on the initial string width and initial loop radii. Using these results, we argue that gravitational wave production is dominated by kinematical instead of geometrical considerations.  

\end{abstract}
\maketitle

\section{Introduction} \label{sect:intro}

  The detection of Gravitational Waves (GW) from black hole (BH) binaries \cite{Abbott:2016blz} by the LIGO/Virgo collaboration marked the start of a new era of observations. Beyond astrophysical objects such as BH and neutron stars, this paved the way for the use of GW to search directly for signatures of new physics. One of the key targets of this search is the existence of a network of cosmic strings \cite{Abbott:2009rr,Aasi:2013vna,Abbott:2017mem,LIGOScientific:2019vic}.

  Cosmologically, cosmic string networks naturally arise after a phase transition in the early universe, possibly during GUT symmetry breaking \cite{Kibble:1976sj,Vilenkin:1981kz,Hindmarsh:1994re,
  Vilenkin:2000jqa,Jeannerot:2003qv,Copeland:2009ga}.  These networks are known to be a source of gravitational waves, and there is a large literature concentrating on the \emph{stochastic}  background of weak field emission of GW through cusps, travelling kinks and kink-kink interactions of the strings \cite{Vilenkin:1981bx,Vachaspati:1984gt,Hogan:1984is,Vachaspati:1987sq,
Garfinkle:1987yw,Garfinkle:1988yi,Quashnock:1990wv,Sakellariadou:1990ne,
Hindmarsh:1990xi,Caldwell:1991jj,Allen:1991bk,Allen:1994iq,Casper:1995ub,Caldwell:1996en,
Allen:2000ia,Damour:2000wa,Damour:2001bk,Damour:2004kw,
Binetruy:2010cc,Ringeval:2017eww,Blanco-Pillado:2017rnf,Jenkins:2018lvb,Drew:2019mzc,Auclair:2019wcv}. This signal is the total integrated power of \emph{incoherent} GW from all such individual emissions, i.e. the sum of all individual emissions which themselves are too weak to be directly detected. Furthermore, these networks may  manifest themselves through other channels, such as their imprints via lensing on the Cosmic Microwave Background \cite{Vilenkin:1984ea,Ade:2013xla}.

Complementarily, one can also search for localized \emph{coherent} events of these strings. Coherent events are those that are individually energetic enough to be detected directly. Such events can occur, for example, when the strings self-interact through the formation of sharp cusps, through the collisions of travelling kinks that are formed during the intercommutation (i.e. collisions) of cosmic strings, or when cosmic string loops collapse. Such a search requires the construction of GW waveform templates -- parameterized coherent time/frequency domain signals which can then be searched via match-filtering in the detector signal stream or identified within a burst search.  We emphasise that searches for stochastic and coherent signals are complementary -- the non-detection/detection of one does not imply the non-detection/detection of the other.

\begin{figure}[t]
\begin{center}
{\includegraphics[width=1.0\columnwidth]{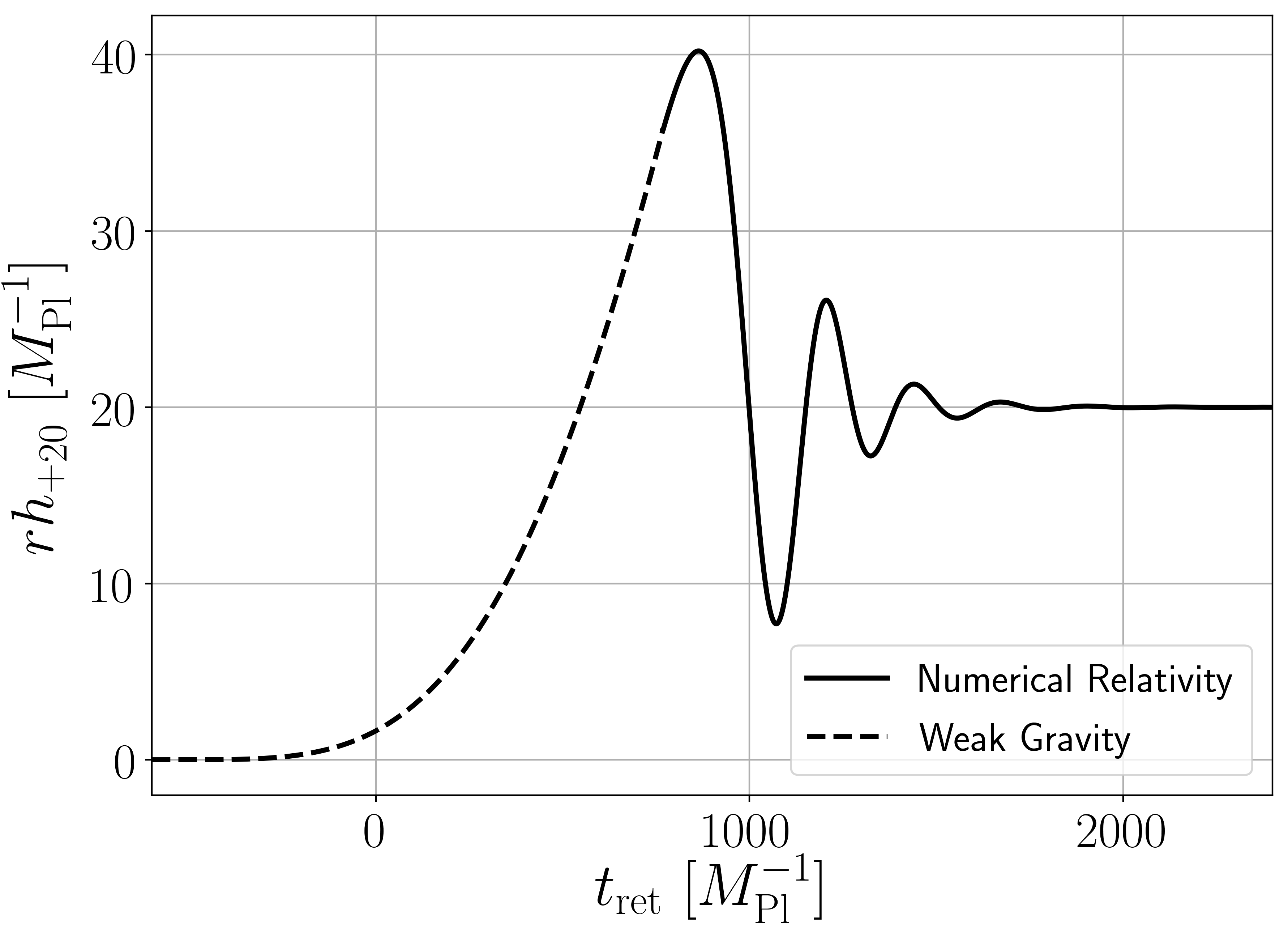}}
  \caption{{\bf Strain waveform}: The $l=2$, $m=0$ strain mode for a cosmic string loop collapse into a black hole with $G\mu = 4\times 10^{-3}$ and $R_0= 600~ \mpl^{-1}$. The dotted signal was calculated using the semi-analytical approach while the solid line is from the integration of the NR signal. The strain exhibits a large memory due to the aspherical loss of matter ejecta during merger, ending with a characteristic ringdown after the black hole is formed.  A summary movie of the simulation can be found \href{https://youtu.be/-dhYA2788LA}{here} \cite{GWmovie}.}
 \label{fig:strainplots}
\end{center}
\end{figure}

  In the literature, collapsing cosmic string loops have been considered as seeds in the formation of primordial black holes \cite{Hogan:1984zb,Hawking:1987bn,PhysRevD.43.1106,Garriga:1993gj,Caldwell:1995fu,
  Cheng_1996,MacGibbon:1997pu,Wichoski:1998ev,Hansen:1999su,Nagasawa2005,Carr:2009jm,Bramberger:2015kua,
  Bertone:2019irm,James-Turner:2019ssu}. Recently, we presented the first investigation of  the collapse of circular loops with full general relativity \cite{PhysRevD.99.104028}\footnote{Work had been done in the past for infinite straight strings and traveling waves in the context of full general relativity \cite{PhysRevD.32.1323,PhysRevLett.59.740,LINET1987240,
  PhysRevD.36.3663,PhysRevD.39.1084,PhysRevD.41.1112,PhysRevD.42.1960}.}. By solving the full non-linear system of Abelian-Higgs field equations coupled to 3+1 Einstein gravity, we showed that the main two outcomes were dispersion and black hole formation. If the loop is not massive enough or thin enough, it will unwind and disperse all the energy into scalar, gauge and gravitational radiation. However, a black hole can be formed, resulting in a large emission of gravitational waves.

\begin{figure*}[t]
\begin{center}
{\includegraphics[width=2\columnwidth]{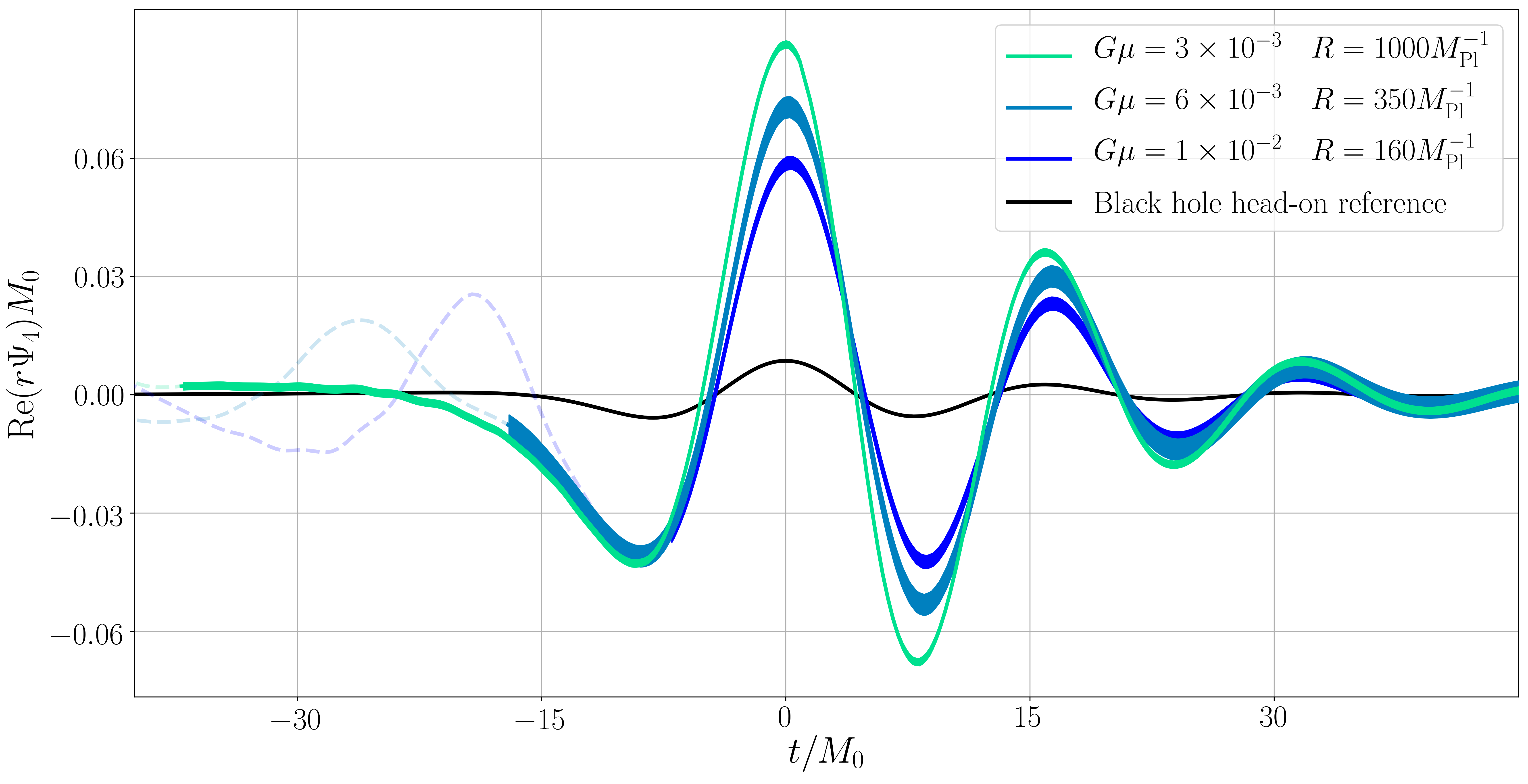}}
    \caption{{\bf Gravitational wave signals as a function of string tension $G\mu$} and black hole head-on reference \cite{Clough:2015sqa}: The signal is normalised with the initial mass of the system and shifted such that the maximum of $r\Psi_4$ coincides at time $t = 0$, for three cases from table (\ref{table:params}) for $G\mu = \{3\times 10^{-3},~6\times 10^{-3},~1\times 10^{-2}\}$ and corresponding mass $M_0 = \{18.85\mpl,~13.19\mpl,~10.05\mpl\}$.  The relationship between $\Psi_4$ and detector strain $h$ is given in \eqref{eq:strainh}.  The thickness of the line is an estimate of the numerical error. Unphysical parts of the signal are de-emphasised using ticked lines with different transparencies.  We find that smaller $G\mu$ have larger amplitudes and hence produce more gravitational wave radiation (with $2.2 \%$ for $G\mu=2\times 10^{-3}$ with $R=1600\mpl^{-1}$). The rest of the initial mass goes into the black hole and matter radiation. A table summary of all the runs is shown in (\ref{table:params}). } 
  \label{fig:GmuRun}
  \end{center}
  \end{figure*}
  
  In this paper, we compute this corresponding coherent GW \emph{strain} in the time-domain -- see fig. \ref{fig:strainplots}. In other words, we compute the GW strain waveform from individual GW events from the collapse  to black holes of cosmic string loops, which is manifestly a strong gravity event. 
 
 We show that the coherent GW strain signals from the collapse of cosmic string loops are dominated by  two major components. The first component is that of a large gravitational wave memory \cite{PhysRevD.45.520,PhysRevLett.67.1486} effect during the merger, generated by a large aspherical ``jet-like'' ejection of matter radiation.  The second component is that of the final ringdown phase post-BH formation, with the initial collapse stage being a subdominant contribution to the total signal.  We also find that the efficiency of GW production is around ${\cal O}(1)\%$ of the total cosmic string mass. This efficiency is  dependent on the cosmic string tension $G\mu$, with \emph{lower} tension producing \emph{more} GW -- up to $2.2\%$ for $G\mu = 2\times 10^{-3}$, which is the lower bound of the parameter space studied in this work. In comparison, the efficiency for head-on BH mergers and inspiral merger is $0.06\%$ and $\sim 5\%$ respectively \cite{PhysRevD.84.084038,PhysRevLett.95.121101}. We will comment on this somewhat counter-intuitive result in section \ref{sect:conclusions}.

  Coherent GW events are categorized by its energy (``loudness'') and its characteristic frequency. The distance $d$ from which one of these events could be observed by current and future GW detectors is given by
  \begin{equation}
  \left(\frac{d}{10~\mathrm{Mpc}}\right)\sim \sqrt{\frac{E_\mathrm{GW}}{M_\odot}}~\left(\frac{10^{-19}}{h}\right)
  \end{equation}
  where $E_\mathrm{GW}$ is the energy emitted in GWs and $h$ is the strain sensitivity of the detector. Roughly speaking, interferometers are optimized to detect GW induced strain of $h\sim 10^{-21}$ around a finite frequency domain -- for the LIGO/Virgo  interferometers this is $f\sim 10-1000$ Hz.  In the case of GW events when a black hole is formed, the quasinormal mode (QNM) frequency of the characteristic ringdown phase is determined by its mass. Combined, this  means that LIGO/Virgo is sensitive to $E_{\mathrm{GW}} \sim M_{\odot}$ events at around 100 Gpc. Thus to produce coherent GW observable by LIGO/Virgo one must produce sufficiently energetic (``loud'') events at its detector frequency\footnote{The signal is redshifted as it travels from the progenitor to the detectors, but this effect is small.}. This means that LIGO/Virgo will be sensitive to cosmic string loop events\footnote{For binary black hole mergers, the efficiency is about $5\%$, i.e. $5\%$ of the merger mass is converted to $E_{\mathrm{GW}}$, putting them into the peak sensitivity window of LIGO/Virgo (${\cal O}(1\sim 100) M_{\odot}$ black holes) as designed.} of around $100M_{\odot}$ at a distance of about 1 Gpc \cite{PhysRevD.99.104028}.

To check the dependency of the waveforms and energy as a function of the initial conditions and parameter of the cosmic string loops, we compute the waveforms for the three main parameters of the system. The first parameter is the string tension $G\mu$ which specifies the underlying theory. The next two parameters, the initial radius $R_0$ and the width of the string $\delta$, define the initial string geometry. We find evidence that the \emph{the mass normalized waveforms depend strongly on the string tension $G\mu$, and weakly on the string width $\delta$ and initial string radii $R_0$}, for the regime $G\mu > 10^{-3}$.   Hence, it follows that the GW production efficiency of collapsing cosmic string loops is only weakly dependent on initial string loop radii $R$ and the width of the string $\delta$ -- at least for the parameter space studied in this work. Combined with the fact that the power is dependent on string tension $G\mu$ -- and this sets the loop velocity at BH formation -- we argue that the generation of GW is driven by collapse kinematics instead of the geometry of the system. 

The paper is organized as follows. In section \ref{sect:numset}, we describe the Abelian Higgs cosmic string model and recap some previous results. In section \ref{sect:paramGW}, we describe the parametric dependences of GW power from both string geometry and string model for cosmic string collapse events. In section \ref{sect:GWForms}, we show how the waveform is not degenerate to other known BH merger processes, and we derive the full coherent time-domain GW strain waveform from a combination of semi-analytic and numerical results. We discuss the prospects and strategies for a direct detection search and conclude in section \ref{sect:conclusions}.

  \section{Abelian Higgs String Loops} \label{sect:numset}

  The action of the Abelian Higgs model minimally coupled to gravity\footnote{We use the $-+++$ convention for the metric, and set $\hbar=c=1$ and $\mpl=1/\sqrt{G}$.} is
  \begin{equation}
  S=S_{EH}-\int d^4x\sqrt{-g}\left[(D_{\mu}{\phi})^{*}(D^{\mu}{\phi})+\frac{1}{4}F_{\mu\nu}F^{\mu\nu}+V(\phi) \right]
  \end{equation}
  where $S_{EH} = \int dx^4 \sqrt{-g}(R/16\pi G)$, $D_{\mu}=(\partial_{\mu}-ieA_{\mu})$ is the covariant derivative  with its $U(1)$ gauge field $A^{\mu}$ with field strength tensor 
  \begin{equation}
  F_{\mu\nu} = \partial_{\mu}A_{\nu}-\partial_{\nu}A_{\mu}~,
  \end{equation}
  and  $V(\phi)$ is the sombrero potential of the complex scalar field $\phi$ 
  \begin{equation}
  V(\phi) = \frac{1}{4}\lambda \left(\left|\phi\right|^2-\eta ^2\right)^2~,
  \end{equation}
  where $\eta$ is the symmetry breaking scale.

  For simplicity, we set the charge $e$ and the dimensionless coupling constant $\lambda$ to obey the critical coupling limit 
  \begin{equation}\label{eqn:critcoup}
  \beta = \frac{\lambda}{2e^2} = 1~,
  \end{equation}
  in which the Higgs and vector masses are identical and the string tension $\mu$ is related to the symmetry breaking scale as
  \begin{equation}\label{eqn:Gmu}
  \mu = 2\pi\eta^2~.
  \end{equation}

  The coupling constant $\lambda$ and the string tension $G\mu$ set the width of string as
  \begin{equation}\label{eq:thickness}
  \delta = \sqrt{\frac{2\pi}{\lambda\mu}}~.
  \end{equation}

  In \cite{PhysRevD.99.104028}, we constructed the initial conditions to a circular cosmic string loop. The mass of such a configuration of radius $R_0$ is given by 
  \begin{equation}\label{eq:mass}
  M_0= 2\pi\mu R_0
  \end{equation}
  which is independent of the coupling constant $\lambda$.

Also in \cite{PhysRevD.99.104028}, we showed that the hoop conjecture argument accurately predicts that an initially static loop with radius $R_0$ and tension $G\mu$ will form a black hole as long as the condition
  \begin{equation}\label{eq:pred}
  R_0>\sqrt{\frac{1}{8\pi\lambda}}\left(G\mu\right)
  ^{-3/2}\mpl^{-1}~,
  \end{equation} 
  is satisfied.

  \section{Parametric dependence of GW signals} \label{sect:paramGW}

  In this section we study how the gravitational wave signal changes when we vary the parameters of the model: the string tension $G\mu$, the initial loop radius $R_0$ and the string width $\delta$. 

  We first focus on the string tension $G\mu$. We performed a series of simulations with the string parameters shown in table (\ref{table:params}) with fixed $\lambda=2$. Since varying $G\mu$ substantially changes the mass of the string (see \eqref{eq:mass}),  for each choice, we choose its initial $R_0$ to ensure that a black hole can be formed (i.e. obey the condition \eqref{eq:pred}). 

In Fig. \ref{fig:GmuRun}, we show the time domain gravitational waveforms in terms of the (mass normalized) $r\Psi_4$ Weyl scalar for the cases\footnote{We show the results of the other simulations in the appendix, Fig. \ref{fig:allsignals}.} of $G\mu = \{3\times 10^{-3},~6\times 10^{-3},~1\times 10^{-2}\}$ with corresponding mass $M_0 = \{18.85,~13.19,~10.05\}\mpl$. For the cases investigated, we find the maximum efficiency is $2.2\%$ for the case of $G\mu = 2\times 10^{-3}$. 

\begin{figure}[t]
\begin{center}
{\includegraphics[width=1.0\columnwidth]{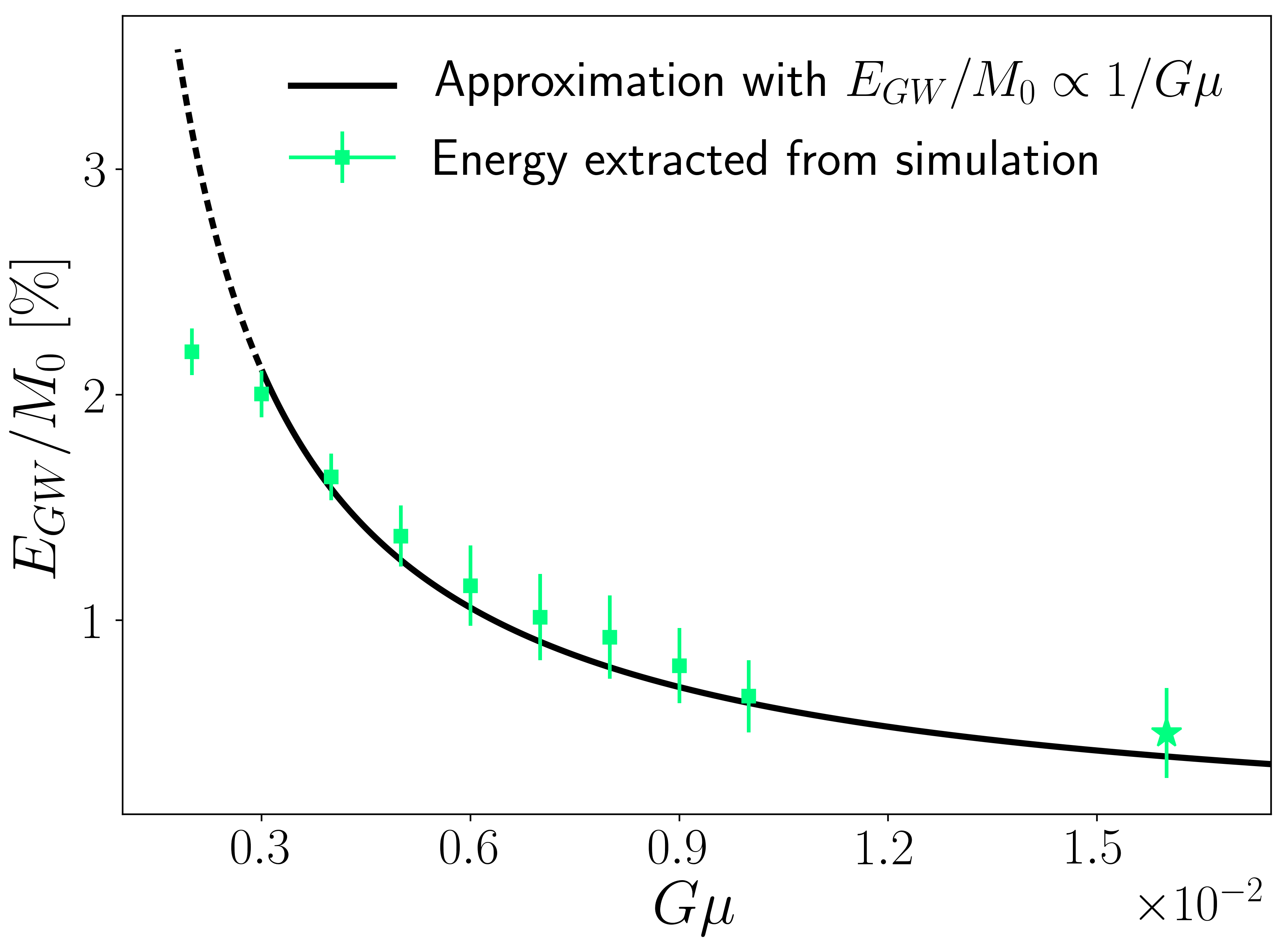}}
\caption{{\bf Efficiency in GW production vs string tension}: We find that the efficiency $E_{\mathrm{GW}}/M_0\propto {\cal A}\left(16\pi^2 G\mu\right)^{-1}$ obey a simple power law with ${\cal A}= 10^{-2}$ (solid line).  The simulation parameters and results are tabulated in Tab. (\ref{table:params}) while the star-dotted point on the right is the result from our previous paper \cite{PhysRevD.99.104028}. Note that the last data point to the left may signal the turnover of the inverse power law $1/G\mu$.}
\label{fig:E_GW}
\end{center}
\end{figure}

The energy radiated in GWs can be estimated from the $r\Psi_4$ Weyl scalar by \eqref{eq:energyPsi4}. The efficiency of GW production normalized over total string mass, $E_{\mathrm{GW}}/M_0$ is shown in Fig. \ref{fig:E_GW}. Interestingly, we find that this scales as 
  \begin{equation}\label{eq:energy}
  \frac{E_\mathrm{GW}}{M_0} =  \frac{\mathcal{A}}{16\pi^2}\frac{1}{G\mu}
  \end{equation}
  where $\mathcal{A}$ is a numerical factor found to be approximately $\mathcal{A}\approx 10^{-2}$. Intriguingly, this means that \emph{smaller tension leads to greater efficiency}, with the caveat that we have only explored a small regime of the total possible parameter space. This scaling clearly cannot be unbounded as $G\mu \rightarrow 0$, and must turnover at some point. We will discuss this further in section \ref{sect:conclusions}.

  We can also explore the dependence of GW emissions as a function of string width $\delta$ and initial radius $R_0$.  In \cite{HAWKING199036}, using purely geometrical arguments, Hawking computed the efficiency of GW emitted from an infinitesimally thin cosmic string loop, and showed that it has an upper bound of $29\%$. This is obtained by assuming that the initial horizon of the black hole is a thin disk, and then computing the difference of the disk's total area with the area of the final Schwarzschild black hole.  Hence, it is plausible that if the initial horizon of the black hole is less disk-like and more spheroidal, the efficiency will become smaller since the initial horizon area will then be greater (and the difference with the area of the Schwarzschild black hole is smaller). To test for this idea, we can define a dimensionless  ``thickness'' parameter,
  \begin{equation}
  \frac{\delta}{R_0}  = \sqrt{\frac{2\pi}{\lambda\mu}}\frac{1}{R_0}\label{eqn:hawking_scaling}~,
  \end{equation}
  such that 
  a cosmic string is ``thin'' if $\delta/R_0$ is small and ''thick'' if $\delta/R_0$ is close to unity. In the infinitesimally thin limit, $\delta/R_0\rightarrow 0$.  Our argument above suggests that the GW efficiency should increase as $\delta/R_0$ decrease, with the Hawking limit being $\delta/R_0=0$. However, as we will show in below, this is not borne out by our numerical simulations, at least in the limited range of parameters we are able to explore.  We test this argument by performing simulations with varying string width $\delta$ and radius $R_0$, while keeping other parameters fixed as follows.

\begin{figure}[t]
\begin{center}
{\includegraphics[width=1.0\columnwidth]{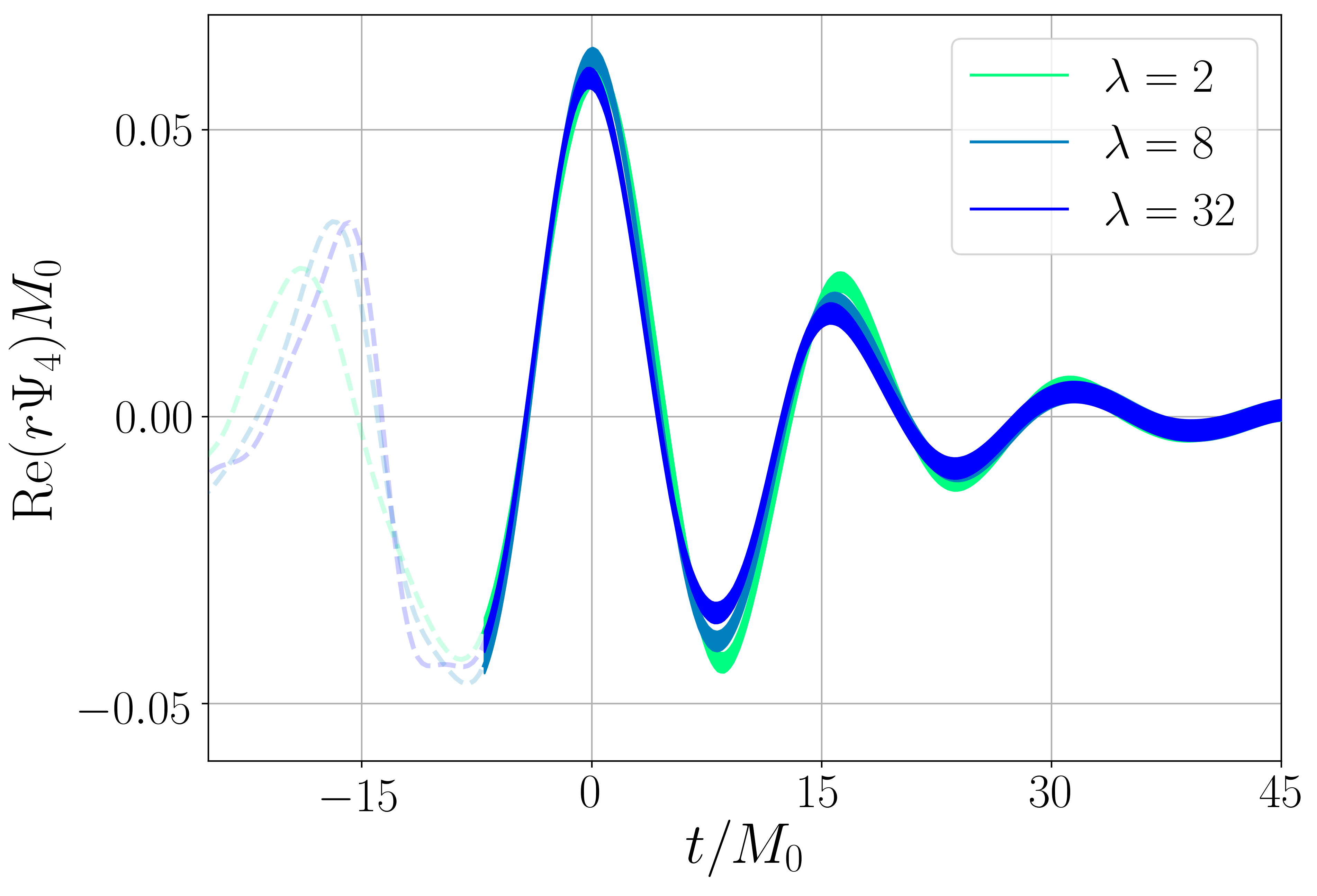}}
  \caption{{\bf Gravitational wave signals for different width $\delta$}: The plot shows the mass normalized Weyl scalar $r\Psi_4$ for $G\mu=1\times 10^{-2}$, $R_0=160\mpl^{-1}$ but with different configurations obtained by varying the string width $\delta$ using expression \eqref{eq:thickness} by half ($\lambda=8$) and quarter ($\lambda=32$). The thickness of the lines indicates the numerical error. This  illustrates that the GW signal does not strongly depend on string width $\delta$.}
 \label{fig:DifferentLambda}
\end{center}
\end{figure}

  \emph{String width $\delta$ dependence}: We performed three simulations with varying $\lambda = \{2,~8,~32\}$ which corresponds to string widths $\delta=\{\delta_2,~\delta_2/2,~\delta_2/4\}$) with $\delta_2=17.72\mpl^{-1}$, while fixing $G\mu=1\times 10^{-2}$ and initial radius $R_0=160\mpl^{-1}$. From the results shown in Fig. \ref{fig:DifferentLambda}, we see that the signals only depend weakly on string width. 

\begin{figure}[t]
\begin{center}
{\includegraphics[width=1.0\columnwidth]{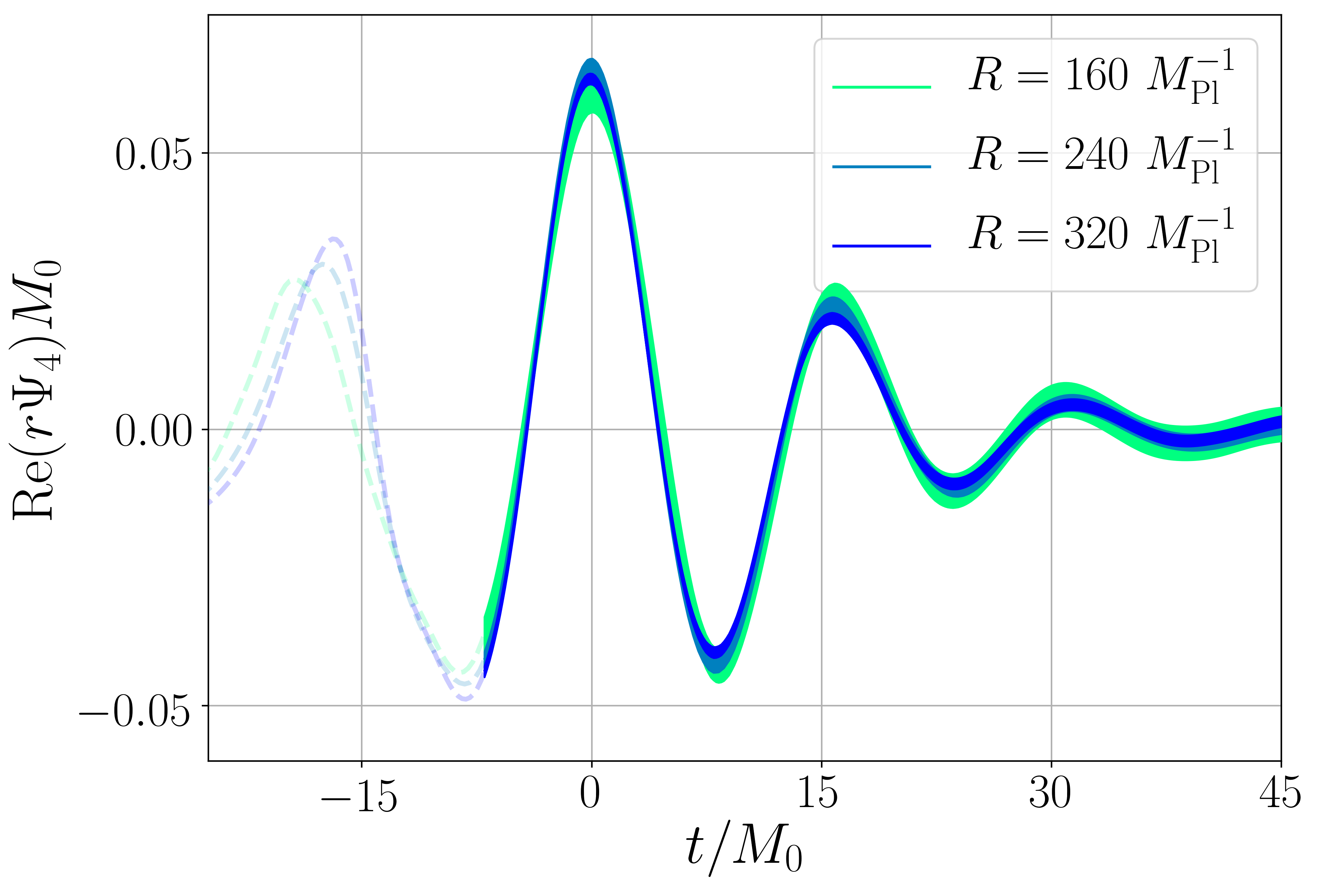}}
  \caption{{\bf Gravitational wave signals for different radii}:  The plot shows the mass normalized Weyl scalar $r\Psi_4$ for the radii $R_0$ $\{ 160,~ 240,~ 320 \}\mpl^{-1}$, with fixed width $\delta=17.72\mpl^{-1}$ and constant tension $G\mu=1\times 10^{-2}$. The thickness of the lines indicates the numerical error. This illustrates that the GW signal does not strongly depend on the string radii.}
 \label{fig:DifferentRadii}
\end{center}
\end{figure}

  \emph{Initial radius $R_0$ dependence }: We performed three simulations with varying $R_0  = \{ 160,~ 240,~ 320 \}\mpl^{-1}$ at fixed $G\mu = 1\times 10^{-2}$  and $\lambda=2$. Since the mass scales with $R_0$ and the ringdown frequency of a black hole is inversely proportional to its mass, we normalise the signal with their initial mass. From the results shown in Fig. \ref{fig:DifferentRadii}, we find that the normalised signal at most scales weakly with $R_0$. 

The above results suggest that the GW emission efficiency is only weakly dependent on initial string dimensionless thickness $\delta/R_0$. 

On the other hand, the numerically obtained scaling \eqref{eq:energy} can be suggestively rewritten as
\begin{equation}
\frac{E_{\mathrm{GW}}}{M_0} = {\cal A}\frac{\gamma(t_{\mathrm{BH}})}{4\pi}~,
\end{equation}
where $\gamma$ is the Lorentz factor of the string infall velocity and $t_{\mathrm{BH}}$ is black hole formation time, i.e.
\begin{equation}
\gamma(t_\mathrm{BH}) = \frac{1}{4\pi G\mu}~. \label{eqn:gammaBH}
\end{equation}

We can derive \eqref{eqn:gammaBH} as follows. In \cite{PhysRevD.99.104028}, we have shown that the dynamics of a radius $R_0$ cosmic string loop during the infall is well described by the Nambu-Goto approximation \cite{Nagasawa:1994md}, for which the position and velocity at some given time are given by
\begin{equation}\label{eq:NGpos}
R(t)=R_0\cos\left(\frac{t}{R_0}\right) ~,~ v_R(t)=\sin\left(\frac{t}{R_0}\right)~.
\end{equation}

The black hole forms approximately when $r_\mathrm{BH}=2GM_0=4\pi R_0 G\mu$, which using \eqref{eq:NGpos} happens at time $t_\mathrm{BH}=R_0\cos^{-1}\left(4\pi G\mu\right)$, so that the velocity at black hole formation is
\begin{equation}\label{eq:BHvel}
v_R(t_\mathrm{BH})=\sqrt{1-16\pi^2\left(G\mu\right)^2}~,
\end{equation}
which using $\gamma = (1-v^2)^{-1/2}$ leads to \eqref{eqn:gammaBH}. For $G\mu=1\times 10^{-2} - 2\times 10^{-3}$, this corresponds to $v(t_\mathrm{BH})\approx 0.9920 - 0.9997$, so it is an ultra-relativistic event.  Note that the velocity equation \eqref{eq:BHvel} does not depend on $\lambda$ and $R_0$.  Physically, the smaller the string tension, for a fixed loop mass $M_0$ the larger the radius of the loop has to be, the longer it takes for the loop to reach the Schwarzschild radius and hence the faster the loop will be moving when the black hole is formed.

Hence we conjecture that the GW emission process is dominated by the kinetic energy of the system, with the string geometry playing only a minor role\footnote{Note that the Hawking argument in \cite{HAWKING199036} assumes that the cosmic string loop is collapsing at the speed of light.}.

\section{Gravitational Strain Waveforms} \label{sect:GWForms}

Our goal in this section is to construct the strain waveform. The gravitational wave strain $h$ as seen by a detector is related to the Weyl scalar $\Psi_4$ by the following equation of motion
\begin{equation}
\ddot{h} = \ddot{h}_{+}+i\ddot{h}_{\times} = \Psi_4~. \label{eq:strainh}
\end{equation}
Thus we would need to integrate \eqref{eq:strainh} to obtain $h$. The details of this integration are described in appendix (\ref{appendix:extension}). 

Furthermore, as we have described in our previous work \cite{PhysRevD.99.104028}, numerically the early time infall signal is contaminated by the presence of unphysical artefacts from the numerical construction of its initial conditions\footnote{These artefacts are generically present for most numerical relativity initial conditions.}. To circumvent this, we note that during this early time period, the infall tracks the trajectory of a Nambu-Goto string until a distance of $\mathcal{O}(\delta)$ \cite{PhysRevD.99.104028}. We use this fact to construct a semi-analytic model of the GW emission during infall  as follows. The modified trajectory is given by
\begin{equation}
R(t) = R_0 \left[\Theta(t_0-t) + \cos\left(\frac{t}{R_0}\right)\Theta(t-t_0)\right]~,
\end{equation}
where the Heaviside functions ensure consistency with the  initial data of our numerical simulations where  the loop is static for $t<t_0$ (see Fig. \ref{fig:spacetime} and Fig. \ref{fig:NGweak}). In Cartesian coordinates $(x,y,z)$ such that $r=\sqrt{x^2+y^2+z^2}$, the stress tensor in the corresponding basis is 
\begin{equation}
T^{\alpha\beta}(t,\mathbf{x}) =  \mu v^{\alpha}v^{\beta} \gamma ~ \delta(r-R(t))\delta(z)~,
\end{equation}
where the velocity is $v^\alpha= (1,~v_R\sin(\phi),~v_R\cos(\phi),~0)$ with
\begin{align}
v_R(t)=\frac{dR}{dt}=\sin\left(\frac{t}{R_0}\right)\Theta(t-t_0)~.
\end{align}

\begin{figure}[t]
\begin{center}
{\includegraphics[width=1.0\columnwidth]{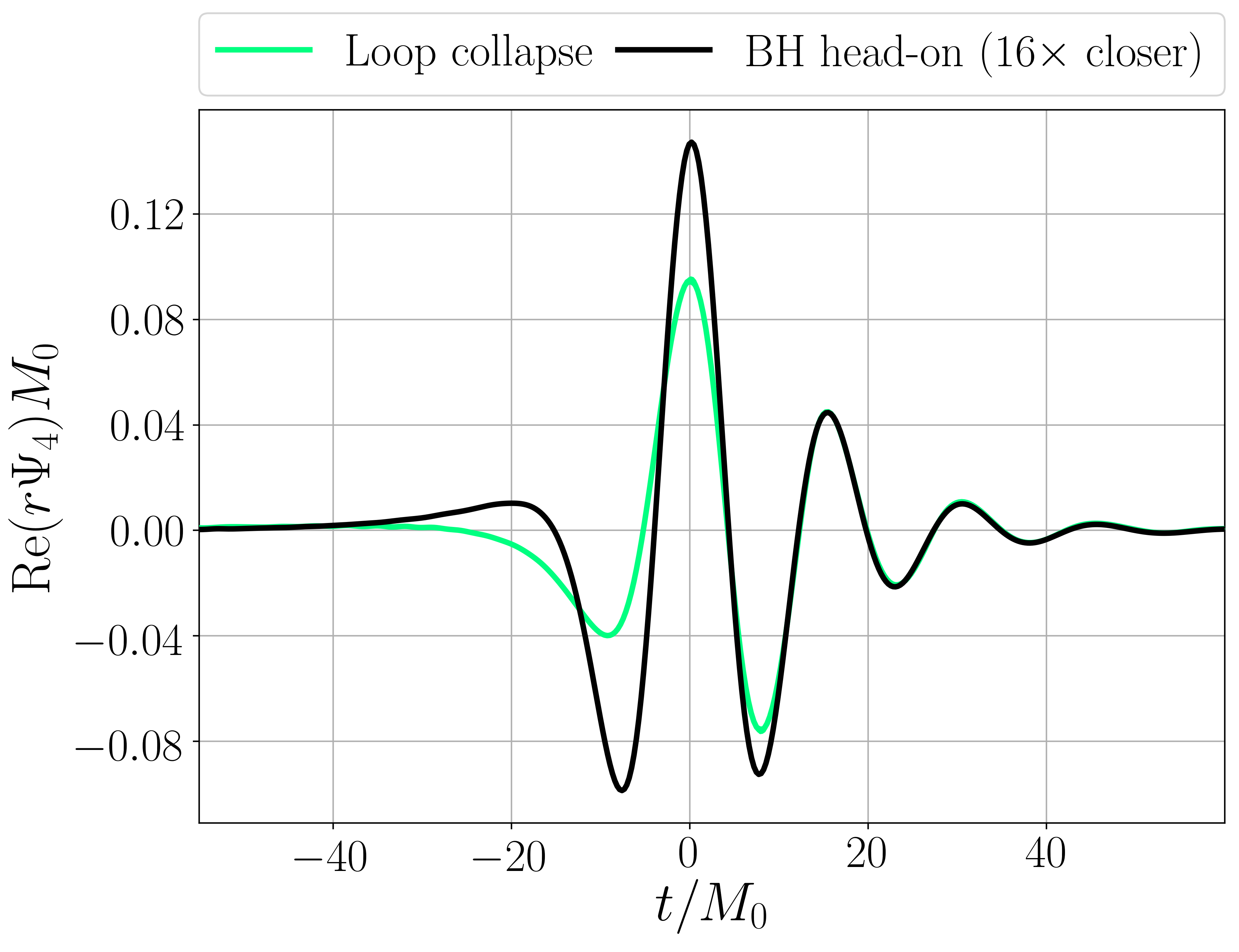}}
  \caption{{\bf String loop and black hole head-on merger comparison}: The $l=2$ $m=0$ strain mode for $G\mu = 2\times 10^{-3}$ with $R_0 = 1600\mpl^{-1}$. Both signals are normalized to mass, but the black hole formed from the head-on collision is $16\times$ closer to the detector. This shows that the signal of the collapse of a cosmic string loop is not degenerate with distance to spin-free BH merger.}
  \label{fig:degenerate}
\end{center}
\end{figure}

The gravitational wave signal of such system is then given in the weak field limit by the standard formula \cite{maggiore2008gravitational}
\begin{equation}\label{eq:fourier2}
r h_{ij}^{TT}(t) = 4G\Lambda_{ij,kl}(\mathbf{n})\int_{-\infty}^\infty \frac{d\omega}{2\pi}\tilde{T}_{kl}(\omega,\omega \mathbf{n}/c)e^{-i\omega t_\mathrm{ret}}
\end{equation}
where $t_\mathrm{ret}=t-r/c$ is the retarded time and is valid for arbitrary velocities, and $\Lambda_{ij,kl}$ is the projector to the traceless-transverse gauge.  The result and details of this calculation for various methods as well as a convergence test can be found in appendix (\ref{appendix:AnalyExtension}) and Fig. \ref{fig:convergencestrain}. We plot the resulting gravitational wave strain for $G\mu=4\times 10^{-3}$ with $R_0=600\mpl^{-1}$ in Fig. \ref{fig:strainplots}.

As one can see,  $r\Delta h_{+} = rh_{+}(\infty)-rh_{+}(-\infty)>0$. This is known as the \textit{gravitational wave memory effect} \cite{Zeldovich:1974gvh,Braginsky1987,PhysRevLett.67.1486, Blanchet:1992br,
PhysRevD.45.520}, which is a large permanent shift in the strain waveform.  The nature of this memory arises from the fact that post-merger, there is a loss of matter emitted axially in an ultra-relativistic jet (Fig. \ref{fig:panel}) -- and hence is highly aspherical --  while its ``incoming'' velocity is zero (i.e. the loop is initially static). This generates a large linear memory shift \cite{Favata:2010zu} akin to that of a core-collapse supernova \cite{Ott:2008wt}.

We can estimate the magnitude of this memory using the linear memory formula \cite{Braginsky1987,PhysRevLett.67.1486}
\begin{equation}
r\Delta h_{ij}^\mathrm{TT}(\theta_A) = \Delta  \sum_A \left[\frac{4GM_A}{\sqrt{1-v_A^2}}\left(\frac{v_A^j v_A^k}{1-v_A\cos\theta_A}\right)^\mathrm{TT}\right]~,\label{eqn:memeffect}
\end{equation}
    where $M_A$ an $v_A$ are the rest mass and asymptotic velocity respectively of ejecta particle  $A$ and $\theta_A$ is the angle between $v_A^i$ and the direction to the detector.  The $\Delta$ expresses the difference between the initial ``incoming'' and ``outgoing'' values. The initial velocity of the loop is $v_A^i=0$. From numerical simulations, it can be seen that the outgoing ejecta is highly beamed like jets in the direction axial to the loop (see Fig. \ref{fig:panel}). In general, to use this formula, one must calculate the flux of ejecta as a function of angle. Since our goal is not to make a precise prediction of its value (we directly obtain this from numerical simulations), but to simply demonstrate that our numerical result is indeed gravitational wave memory, we approximate its magnitude as follows. We assume that all the ejecta is travelling at a constant velocity axially (i.e. perpendicular to the plane of the loop) at $v_A^i = (0,0,\pm v_z)$ where $v_z \sim 1$ (the exact value does not affect the final answer significantly).

We express the right hand side of \eqn{eqn:memeffect} onto a spherical basis at radius $r$ by first rotating each instance of the metric $r \Delta h_{ij} \rightarrow  r \Delta h_{i'j'}(\theta,\phi)$ where $(\theta,\phi)$ are the coordinates on the sphere. We then project the metric onto their traceless and transverse components to obtain
\begin{equation}
\Delta h_{i'j'}^{TT}(\theta,\phi) = \threedsquaremat{\Delta h^+}{\Delta h^\times}{0}{\Delta h^\times}{-\Delta h^+}{0}{0}{0}{0}~,
\end{equation}
where it can be shown that
\begin{equation}
r\Delta h^+  = 2GE_{\mathrm{total}}\frac{v_z^2\sin^2 \theta}{v_z^2\cos^2 \theta-1}~,\qquad r\Delta h^{\times} = 0~. \label{eqn:hplusmem}
\end{equation}
   and $E_\mathrm{total}\approx M_0-M_{\mathrm{BH}} = 1.32 \mpl$ (see Tab. (\ref{table:params})) is the total integrated relativistic flux energy for both matter and GW we directly measured from our simulations. To compare this to our numerical result in Fig. \ref{fig:strainplots}, we project \eqref{eqn:hplusmem} onto the $l=2$, $m=0$ mode as
\begin{equation}
r\Delta h^{+}_{2,0} = \int d\Omega~ r \Delta h^+({}_{-2}Y^{2}_{0})^* \approx ~8~\mpl^{-1}~, \label{eqn:memestimate}
\end{equation}
which about a factor of 2 smaller when compared to the numerical value we obtained, but at the right order of magnitude. We emphasise that \eqref{eqn:memestimate} is just an estimate of the memory assuming the interactions stay within the linear regime, and hence it is not surprising that the true memory is larger.  

\section{Discussion and Prospects for Detection} \label{sect:conclusions}
In this work, we showed that GW production of cosmic string loops that collapse and form black holes scales as
\begin{equation}
\frac{E_\mathrm{GW}}{M_0} =  \frac{\mathcal{A}}{16\pi^2}\frac{1}{G\mu}~,~{\cal A}\approx 10^{-2}~, \label{eqn:Escale}
\end{equation} 
but depends weakly on its initial string width and loop radius. We argue that this strongly suggests that the GW production in such a collapse is dominated by kinematic processes, and not geometric ones. 

Clearly, since $G\mu$ is theoretically not bounded from below, \eqref{eqn:Escale} cannot scale without bound to smaller values as  it violates the  Hawking bound $E_\mathrm{GW}/M_0\rightarrow 0.29$ at $G\mu \approx 2 \times 10^{-5}$. This suggests that there must exist some new scale where this turnover from the inverse power law to some other relationship. This turnover may already be hinted in Fig. \ref{fig:E_GW}, where the $G\mu=2\times 10^{-3}$  point is diverging from expression \eqref{eqn:Escale}, and will be a focus of our future investigations. Furthermore, our cosmic string loops are Planckian in their masses. To generate loops of solar masses require that the loops have large radii -- for example for $G\mu \approx 10^{-10}$ require a loop of around $100$ a.u. \footnote{By equation \eqref{eqn:gammaBH}, such loops would experience a $\gamma(t_\mathrm{BH})\approx \mathcal{O}(10^9)$ Lorentz contraction. In order to numerically simulate this regime, one requires substantial investment of numerical resources and engineering, but in principle can be done with judicious use of the $S_1$ symmetry of the loop.}.

Observations of the CMB \cite{Ade:2013xla} and the LIGO/Virgo search for stochastic GW \cite{Abbott:2017mem,LIGOScientific:2019vic} constraints the current cosmic string tension to $G\mu \lesssim 10^{-14} - 10^{-7}$ -- this value is dependent on the details of the cosmic strings network evolution which is uncertain (and model dependent) \cite{Ringeval:2005kr,Siemens:2006vk,Lorenz:2010sm,BlancoPillado:2011dq, Blanco-Pillado:2013qja,Blanco-Pillado:2019vcs,Lizarraga:2016onn}. This regime is obviously beyond the validity of our scaling argument. While we have only explored a small regime of the possible parameter space and  the amplitude of the GW signal may differ for other parameters, we do not expect the \emph{form} of the GW strain signal shown in Fig. \ref{fig:strainplots} to differ substantially at lower $G\mu$. We also emphasise that strongly gravitating strings such as fundamental strings with $G\mu \sim 10^{-2}$ can also be produced in many popular brane inflation models \cite{Witten:1985fp,Jones:2002cv,Jones:2003da,Polchinski:2004ia}.   Modulo such theoretical concerns about the probability distribution of such events which can only be estimated from large network simulations, we take the agnostic view that their existence can be put into observational test. 

On the other hand, we believe that the large gravitational wave memory of these events is a robust result regardless of the string parameters, since it is sourced by the large aspherical emission of post-collapse debris which we expect to occur regardless.  While GW memory are historically removed from both the detector data streams and theoretical predictions, there is now increasing interest in their search \cite{Divakarla:2019zjj,Boersma:2020gxx} and is currently a goal of the LIGO/Virgo collaboration \cite{Hubner:2019sly}.
 
Both such short signals with little GW production during the infall phase suggests that this it is best looked for in the transient short-during burst channel \cite{Abadie:2012rq,Abbott:2016ezn,TheLIGOScientific:2016uux,Aasi:2013wya,Abbott:2019prv}. This channel makes only minimal assumptions on the expected signal waveform, at the cost of reduced sensitivity to weaker signals. 
One may wonder whether the string loop  burst waveform is degenerate with other processes such as very massive binary black hole inspiral or head-on mergers -- and hence can be picked up by already existing match-filtered searches. The former case is trivial since the lack of an oscillatory pre-merger signal and the fact that the black hole formed the collapse has no spin, are sufficient features to distinguish from a binary black hole inspiral system, and thus it is not degenerate.

For a more symmetric scenario such as a head-on BH-BH merger, in Fig. (\ref{fig:degenerate}) we show that it is not degenerate. While the ringdown signal from the black hole formed from a loop is degenerate with a black hole with the same mass formed from a head-on merger $16\times$ closer, the pre-merger and the merger itself differ considerably. Therefore, it will be distinguishable as long as one has access to the full waveform. 

To detect such weaker signals, one would need to make use of the full match-filtering search, which requires the construction of a parameterised GW waveform template. In this work, we argue that the primary parameter for the construction of such waveform templates is the string tension $G\mu$, with secondary parameters being the initial string width and radii. We undertook the first  steps in the construction of the GW strain waveform  template (Fig. \ref{fig:strainplots}). In an upcoming publication, we will complete the construction of these templates, and use them to search for cosmic string loop collapse signatures in the LIGO/Virgo data stream.

\acknowledgments
We acknowledge useful conversations with Jose Juan Blanco-Pillado, Katy Clough, Ed Copeland, Tim Dietrich, Amelia Drew, Tanja Hinderer, Alex Jenkins, Sebastian Khan, Samaya Nissanke, Paul Shellard, Kepa Sousa, and Andrew Williamson. We would like to thank the Lorentz Center, organisers and participants of the ``Cosmic Topological Defects: Dynamics and Multi-messenger Signatures" workshop. We would also like to thank the GRChombo team
\href{http://www.grchombo.org}{(http://www.grchombo.org/)} and the
COSMOS team at DAMTP, Cambridge University for
their ongoing technical support.  

EAL is supported by an  STFC  AGP-AT  grant  (ST/P000606/1). TH is supported by NSF Grants No. PHY-1912550 and AST-1841358, NASA ATP Grants No. 17-ATP17-0225 and 19-ATP19-0051, and NSF-XSEDE Grant No. PHY-090003. This work has received funding from the European Union’s Horizon 2020 research and innovation programme under the Marie Skłodowska-Curie grant agreement No. 690904. The authors would like to acknowledge networking support by the GWverse COST Action CA16104, ''Black holes, gravitational waves and fundamental physics''.This  work  was  performed  using  the Cambridge  Service  for  Data  Driven  Discovery  (CSD3), part of which is operated by the University of Cambridge Research Computing on behalf of the STFC DiRAC HPC Facility  (www.dirac.ac.uk) and on Leibnitz Supercomputing Center SuperMUC-NG under PRACE grant Tier-0 Proposal 2018194669. The  DiRAC  component  of CSD3 was funded by BEIS capital funding via STFC capital grants ST/P002307/1 and ST/R002452/1 and STFC operations grant ST/R00689X/1.  DiRAC is part of the National e-Infrastructure.

\bibliography{mybib}

\clearpage
\appendix

 
 \section{Extending the waveform}\label{appendix:extension}

\subsection{Integrating the $r\Psi_4$}\label{appendix:GettingStrain}

The GW strain can be obtained directly from integrating the numerically obtained Weyl Scalar $\Psi_4$, 
\begin{equation}
\ddot{h} = \ddot{h}_{+}+i\ddot{h}_{\times} = \Psi_4,
\end{equation}
with the boundary conditions that the emission in gravitational wave power stops at large times and $P_{GW} \propto \dot{h}$ 
\begin{equation}
\lim_{t \rightarrow\infty} \dot{h} =  0~.
\end{equation}
We hence have the freedom to shift $h$
\begin{equation}
h =  h_\mathrm{num} + \Delta h~,
\end{equation}
where $h_\mathrm{num}$ is the gravitational wave strain calculated using a numerical integration technique from $\Psi_4$. However, we found in the simulations that the quasi-normal modes become unreliable after a certain time due to numerical resolution (see Fig. \ref{fig:QNMl2m0fit} for $t \gsim 4500$), which causes substantial errors in the integration. To deal with this, this we substitute the signal with analytical QNMs \cite{Berti:2005ys} for the corresponding $l = 2$ mode. We performed convergence checks in resolution, courant-factor, box-radius and extraction radius, to ensure that all our numerical integrations are converged. 

\subsection{Weak-field gravity extension}\label{appendix:AnalyExtension}
\begin{figure}[t]
\begin{center}
{\vspace{0.77cm}
\includegraphics[width=1.0\columnwidth]{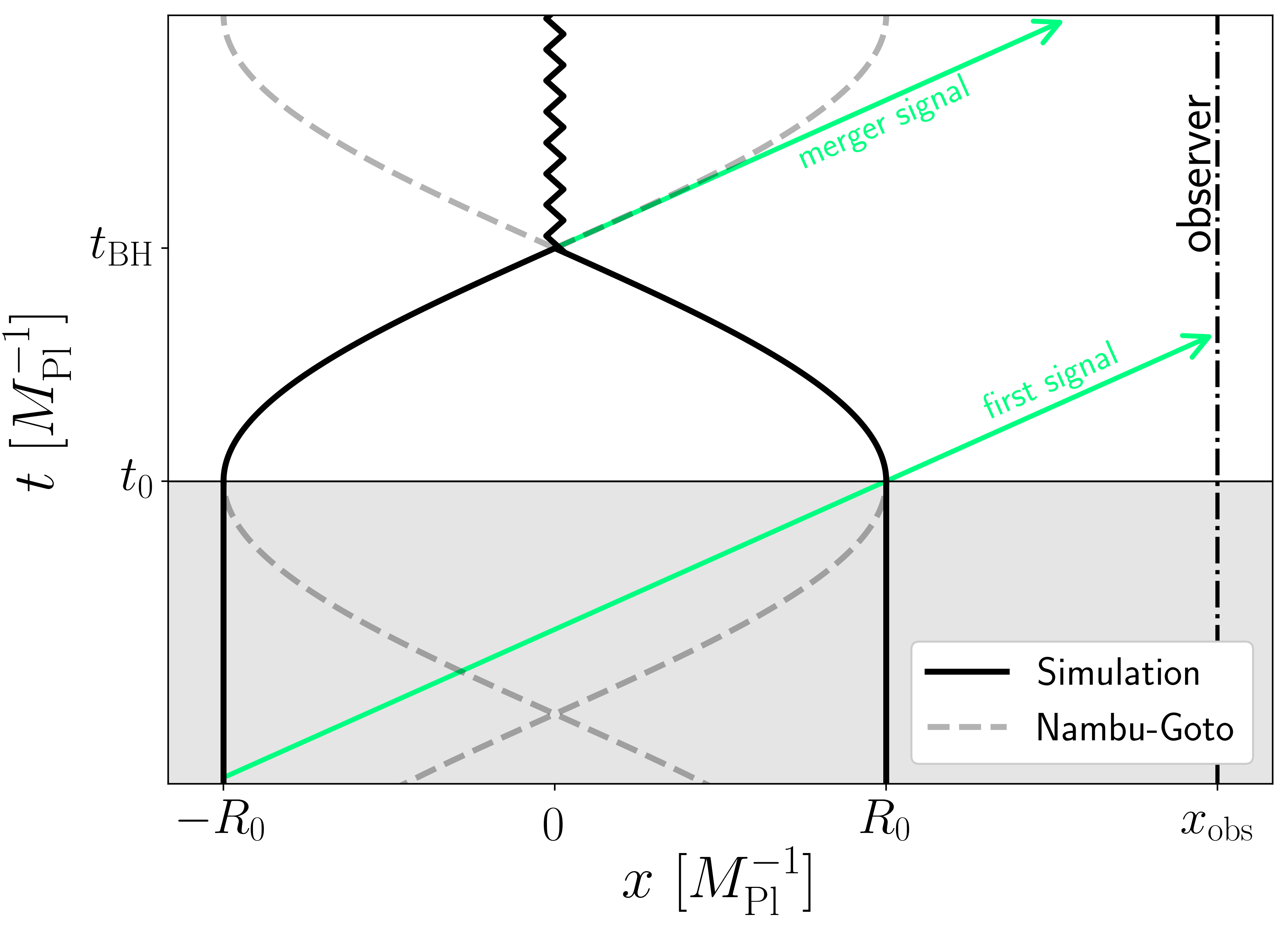}}
  \caption{{\bf Spacetime diagram of loop evolution}: The solid black line represents the loop evolution of our simulations. The loop is initially at rest with radius $R_0$, then starts to collapse at $t_0$ and forms a black hole at $t_\mathrm{BH}$. The dashed grey is the solution of an oscillating loop following the Nambu-Goto action. As shown in Fig. \ref{fig:NGweak}, the first signal an observer at $x_\mathrm{obs}$ receives depends on the past history of the loop (grey shaded area). For the Nambu-Goto case, one would get gravitational radiation coming from the expansion phase of the loop (after it has shrunk to a point in the previous cycle). We cut this spurious signal off by imposing a Heaviside function in \eqref{eq:NGpos}.}
 \label{fig:spacetime}
\end{center}
\end{figure}

\begin{figure}[t]
\begin{center}
{\vspace{0.77cm}
\includegraphics[width=1.0\columnwidth]{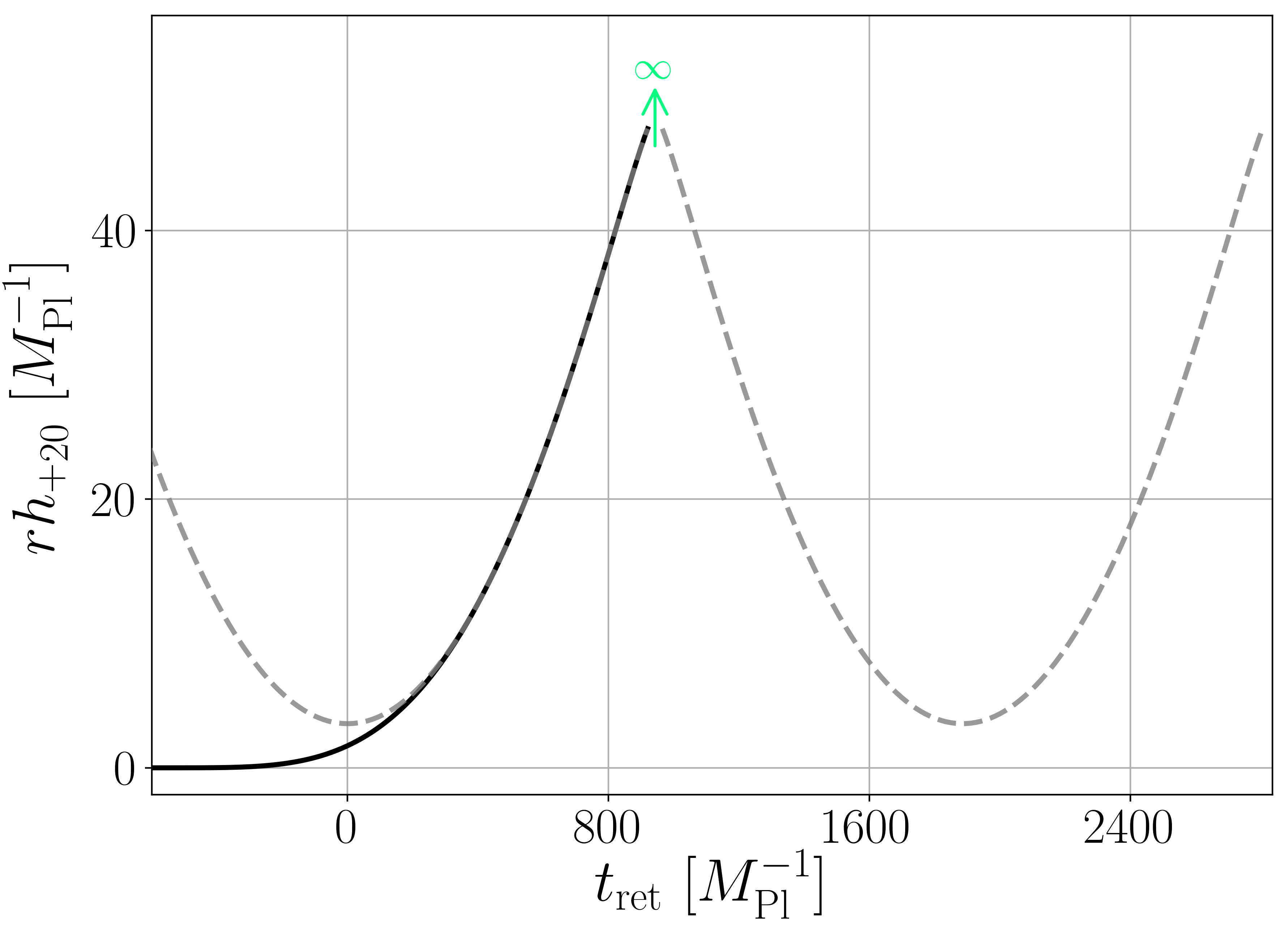}}
	\caption{{\bf GW signal from weak field gravity} for the infall of a loop simulated in this paper (solid black line) and an oscillatory Nambu-Goto loop (dashed grey line). The signal of the former starts at $rh=0$ while for the latter, the observer gets contribution from the expanding regime $(t<0)$ of the Nambu-Goto loop. The weak gravity calculation breaks down when the loop collapses to a point. }
 \label{fig:NGweak}
\end{center}
\end{figure}

To construct the infall signal, we will calculate ithe strain of a collapsing circular and planar cosmic string loop with energy momentum tensor given by
\begin{equation}
T^{\alpha\beta}(t,\mathbf{x}) =  \mu v^{\alpha}v^{\beta} \gamma ~ \delta(r-R(t))\delta(z)
\end{equation}
where we define $r = \sqrt{x^2+y^2}$ and the behaviour of the pre-merger collapse in the weak-field limit is well described by
\begin{equation}
R(t) = R_0 \left[\Theta(t_0-t) + \cos\left(\frac{t}{R_0}\right)\Theta(t-t_0)\right].
\end{equation}
so that $v^\alpha= (1,~v_R\sin(\phi),~v_R\cos(\phi),~0)$ with
\begin{align}
v_R(t)=\frac{dR}{dt}=&\sin(t/R_0)\Theta(t-t_0)~.
\end{align}
where $\delta(t-t_0)$ is the Dirac delta and we set the starting time $t_0=0$ to be consistent with the simulations. Note that we have use the Heaviside Theta functions to impose the initial of the cosmic loop such that it is infinitely static from $t<t_0$, consistent with the initial conditions of our numerical simulations. This is important as the Nambu-Goto loop is oscillating, and hence will contribute GW in the regime $t<t_0$, in contradiction to our numerical simulations (see Figs. \ref{fig:spacetime} and \ref{fig:NGweak}).

The effective GW generated for sources that are relativistic is given by \cite{maggiore2008gravitational}
\begin{equation}\label{eq:fourier}
{r}h_{ij}^{TT}(t) = 4G\Lambda_{ij,kl}(\mathbf{n})\int_{-\infty}^\infty \frac{d\omega}{2\pi}\tilde{T}_{kl}(\omega,\omega \mathbf{n}/c)e^{-i\omega(t-r/c)} ~,
\end{equation}
where $\mathbf{n}$ is the direction of the observer 
\begin{equation}
\mathbf{n} = \left(\sin\theta \sin\phi,\sin\theta\cos\phi,\cos\theta \right)~,
\end{equation}
and  $\Lambda_{ij,kl}(\mathbf{n})$ is the projector to the TT gauge,
\begin{equation}
\Lambda_{ij,kl}(\mathbf{n}) = P_{ik}P_{jl}- \frac{1}{2}P_{ij}P_{kl}~,
\end{equation}
where 
\begin{equation}
P(\mathbf{n}) = \delta_{ij}-n_in_j~.
\end{equation}
\begin{figure}[t]
\begin{center}
{\includegraphics[width=1.0\columnwidth]{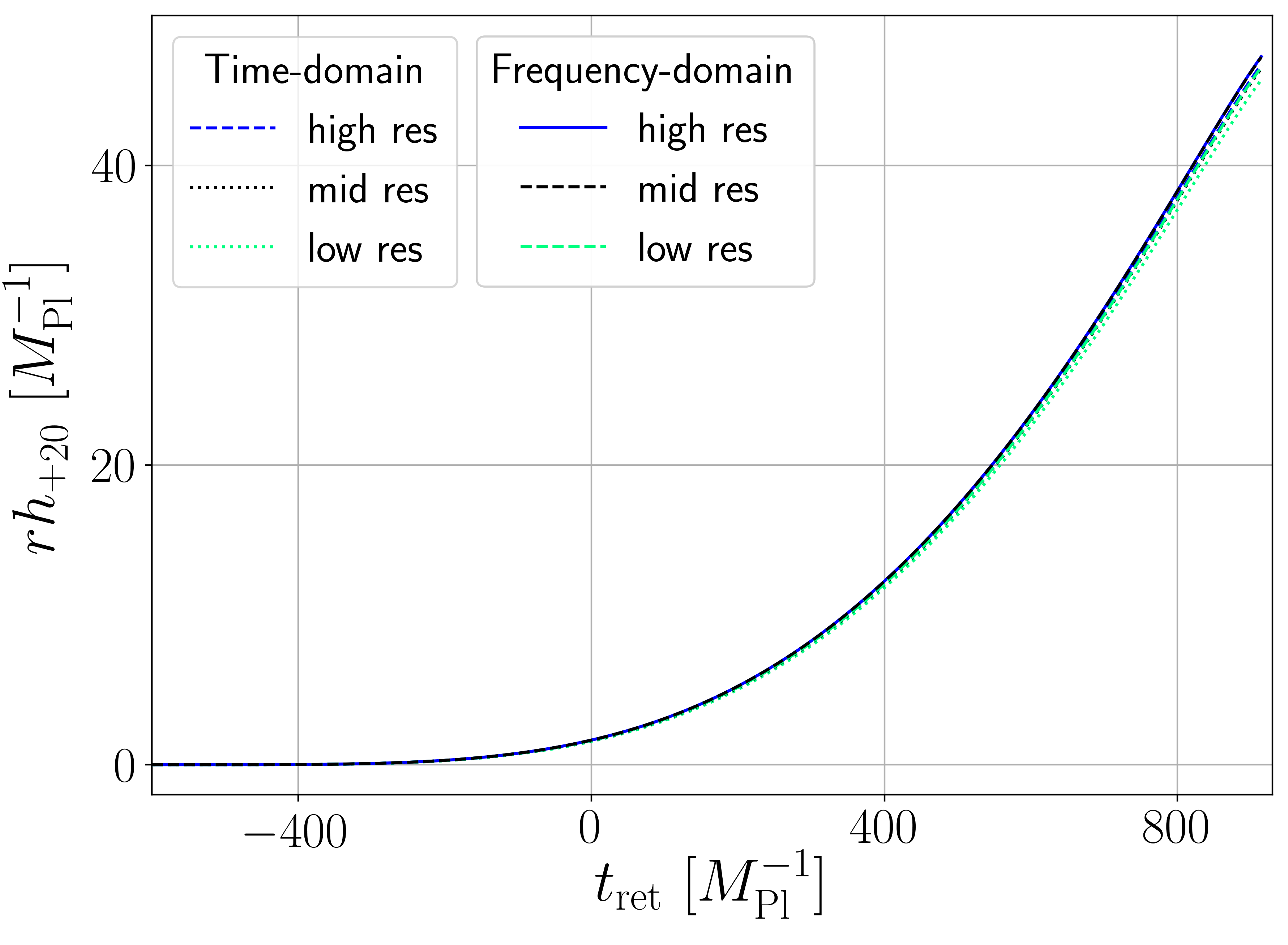}}
  \caption{{\bf Consistency test} between the frequency-domain and time-domain methods for $G\mu=4\times 10^{-3}$ and $R_0=600\mpl^{-1}$. We run both methods with three resolutions, which we refer as low, mid and high. The difference between them becomes smaller as the resolution is increased, indicating that our integration has converged. Both methods recover the same signal. }
 \label{fig:convergencestrain}
\end{center}
\end{figure}
We define the fourier transform as
\begin{equation}\label{eq:emtensor_fourier}
\tilde{T}_{kl}(\omega,\mathbf{k}) = \int {d^4x}~ T_{kl}(t,\mathbf{x}) e^{i\omega t - i \mathbf{k}\cdot\mathbf{x}}~.
\end{equation}
To check the calculation we also calculate the same expression in the time-domain, 
\begin{equation}\label{eq:time}
{r}h_{ij}^{TT}(t) = 4G\Lambda_{ij,kl}(\mathbf{n})\int_{-\infty}^\infty d^3 x~T_{kl}\left(t-\frac{r}{c}+\mathbf{x}\cdot\mathbf{n},\mathbf{x} \right)~,
\end{equation}
we indeed find that both formulations converge to the same result (see Fig. \ref{fig:convergencestrain}).
\subsubsection{Frequency-domain}
We simplify \eqref{eq:fourier} into 
\begin{equation}\label{eq:Integrals}
\begin{split}
rh_{+}(t,\theta,\phi) &= \frac{1}{2} \left( I_{2}(t,\theta )-\cos ^2(\theta ) (I_{1}(t,\theta)-I_{2}(t,\theta))\right), \\
rh_{\times} &= 0
\end{split}
\end{equation}
where the two integrals are 
\begin{equation}
I_{1}(t,\theta)=8 G\mu R_0 \int_{t_1}^{t_2} \frac{\sin(t^\prime/R_0)^2 \Theta(t^\prime)}{\sqrt{A^2-(t^\prime-t_\mathrm{ret})^2}} dt^\prime~,
\end{equation}
and
\begin{equation}
I_{2}(t,\theta)=8 G\mu R_0 \int_{t_1}^{t_2} \frac{\sin(t^\prime/R_0)^2\sqrt{A^2-(t^\prime-t_\mathrm{ret})^2}\Theta(t^\prime)}{A^2}dt'
\end{equation}
with $A=R_0 \left[\Theta(-t^\prime) + \cos\left(t^\prime/R_0\right)\Theta(t^\prime)\right] \sin (\theta)$ and $t_\mathrm{ret}=t-r/c$ the retarded time. These are integrated numerically from $t_1(t,\theta)$ to $t_2(t,\theta)$, defined so that the square root above is well defined. To find these two points, one has to find the roots in $t^{\prime}$ of $R_0^2 \left[\Theta(-t^\prime) + \cos\left(t^\prime/R_0\right)\Theta(t^\prime)\right]^2 \sin (\theta)^2-(t^\prime-t_\mathrm{ret})^2=0$, which we did using a non-linear  numerical solver for every $t$ and $\theta$.

\subsubsection{Time-domain}
Similarly as in \eqref{eq:Integrals}, we can simplify \eqref{eq:time} into 
\begin{equation}\label{eq:rotated_metric}
\begin{split}
rh_{+}(t,\theta,\phi) &= \frac{1}{2} \left( rI_{xx}(t,\theta )-\cos ^2(\theta ) rI_{yy}(t,\theta )\right) ~,~\mathrm{and}~, \\
rh_{\times} &= 0
\end{split}
\end{equation}
where the integrals are
\begin{equation}
I_{xx}(t,\theta) =4 G\mu R_0 \int_{0}^{2\pi}  d\phi^\prime ~\frac{ B^2 \sin^2 (\phi^\prime)}{1  + \cos(\phi^\prime)\sin(\theta) B  },
\end{equation}
and
\begin{equation}
I_{yy}(t,\theta) =4 G\mu R_0 \int_{0}^{2\pi}  d\phi^\prime ~\frac{ B^2 \cos^2 (\phi^\prime)}{1  + \cos(\phi^\prime)\sin(\theta) B  }
\end{equation}
with 
\begin{equation}
\begin{split}
B =\sin\left(\frac{t_\mathrm{ret}+r(\phi^\prime,\theta,t)\cos(\phi^\prime)\sin(\theta)}{R_0}\right)\\ \times~\Theta(t_\mathrm{ret}+r(\phi^\prime,\theta,t)\cos(\phi^\prime)\sin(\theta))~,
\end{split}
\end{equation} 
where one has to first obtain  $r(\phi^\prime,\theta,t)$ by solving 
\begin{equation}
\begin{split}
r-R_0\cos\left(\frac{t_\mathrm{ret}+r(\phi^\prime,\theta,t)\cos(\phi^\prime)\sin(\theta)}{R_0}\right)\\
\times\Theta(t_\mathrm{ret}+r(\phi^\prime,\theta,t)\cos(\phi^\prime)\sin(\theta))=0,
\end{split}
\end{equation}
using a non-linear solver similarly to the frequency approach for $t_1(t,\theta)$ and $t_2(t,\theta)$. However, we need to solve for an additional variable this method, it is numerically much more expensive but we use it to check consistency between both methods, Fig. \ref{fig:convergencestrain}.


\begin{figure}[t]
\begin{center}
{\includegraphics[width=1.0\columnwidth]{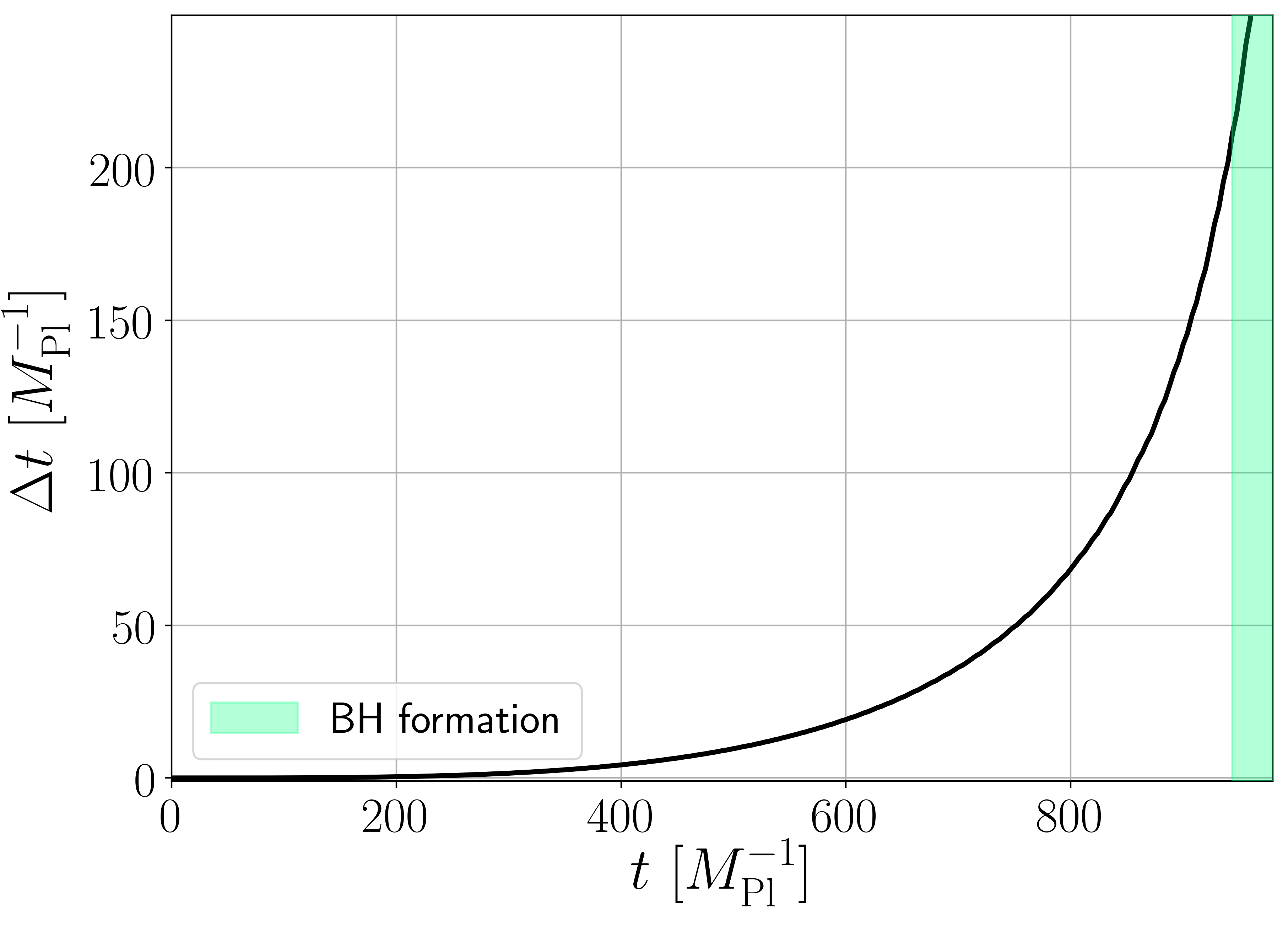}}
\vspace{-1.5em} \caption{\textbf{Time delay} (\eqref{eq:delay}) caused by the dynamical gauge for the case $G\mu=4\times 10^{-3}$, $R_0=600\mpl^{-1}$. We estimate GWs emitted near BH formation to be reaching our extraction radius with $\Delta t\approx 200\mpl^{-1}$ delay in simulation time.}
\label{fig:delay}
\end{center}
\end{figure}

\begin{figure}[t]
\begin{center}
{\vspace{0.77cm}
\includegraphics[width=1.0\columnwidth]{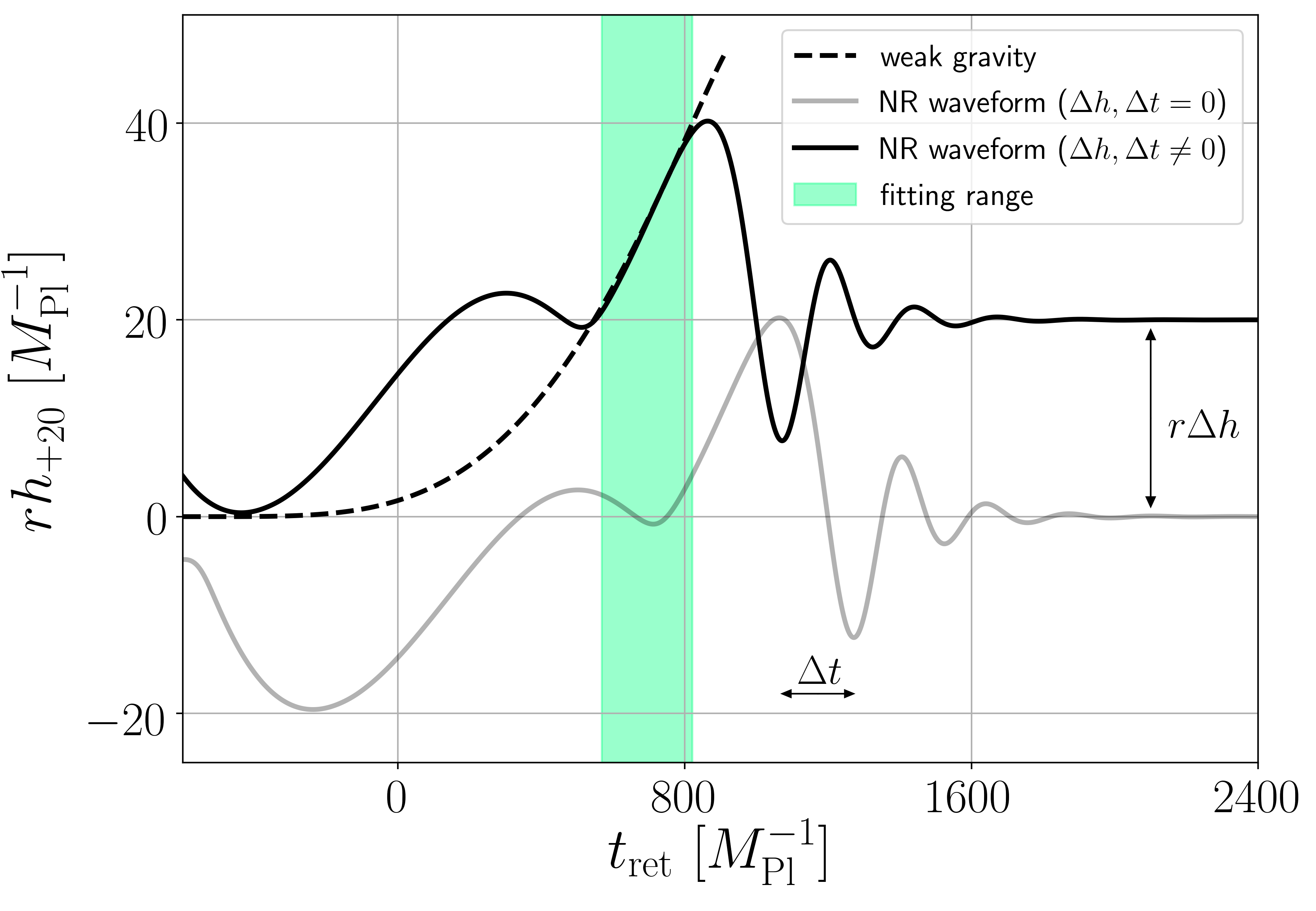}}
	\caption{{\bf Matching the numerical and analytical signals} for the $G\mu=4\times 10^{-3}$, $R_0=600\mpl^{-1}$ case. As estimated in Fig. \ref{fig:delay}, we correct the time delay by shifting the numerical signal by $-\Delta t =-200\mpl^{-1}$. The shaded region indicates where the best fit is being calculated to determine the free shift $r\Delta h$, which is found to be $r\Delta h \approx 20\mpl$. }
 \label{fig:fitting}
\end{center}
\end{figure}

\subsection{Fitting to the NR signal}\label{appendix:fittingtoNR}

\begin{figure}[t]
\begin{center}
{\includegraphics[width=1.0\columnwidth]{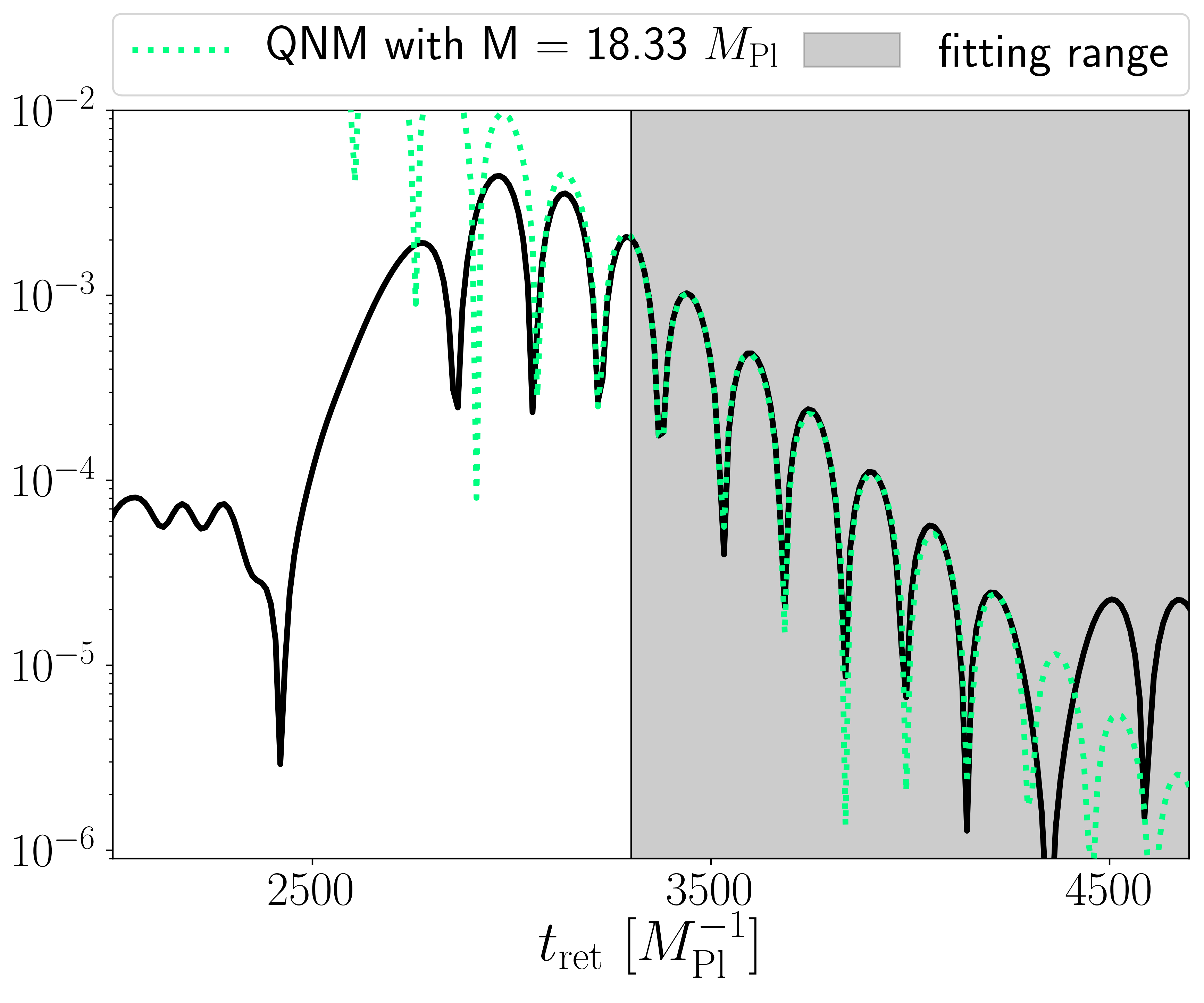}}
\vspace{-1.5em} \caption{\textbf{Fitting $l = 2$ $ m = 0$ Quasinormal mode:} We cut off the numerical  signal at $t= 3300\mpl^{-1}$ and search for the mass that best fits the analytic quasi-normal mode with the signal. We find a good fit with the mass $18.33\mpl$ for $G\mu=2\times 10^{-3}$ and $R_0=1600\mpl^{-1}$.  }
\label{fig:QNMl2m0fit}
\end{center}
\end{figure}

\begin{figure}[t]
\begin{center}
{\includegraphics[width=1.0\columnwidth]{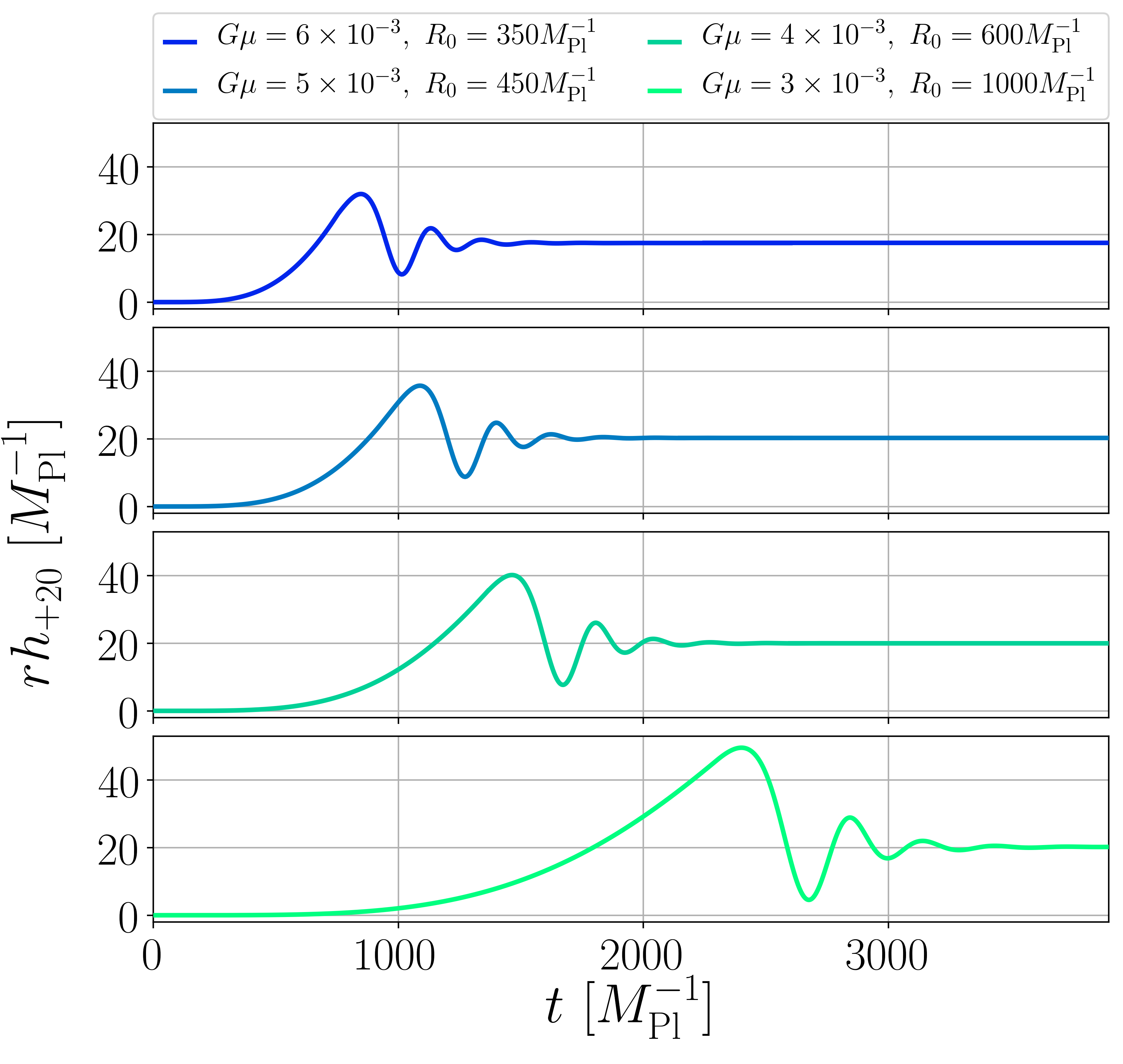}}
\vspace{-1.5em} \caption{\textbf{Gravitational waveforms} for $(G\mu,R_0)=\lbrace(6\times 10^{-3},350\mpl^{-1}),(5\times 10^{-3},450\mpl^{-1}),(4\times 10^{-3},600\mpl^{-1}),(3\times 10^{-3},1000\mpl^{-1})\rbrace$. The numerical signals have been corrected the delays $\Delta t=\lbrace 160\mpl^{-1},~180\mpl^{-1},~200\mpl^{-1},~300\mpl^{-1}\rbrace$ respectively, estimated via \eqref{eq:delay}. The figure shows how larger loops have a longer infall and the memory is about the same for the last three cases, which is expected since the total radiation in GWs and matter is very similar $M_0-M_{\mathrm{BH}}\approx 1.25\mpl$, while for $(G\mu,R_0)=(6\times 10^{-3},350\mpl^{-1})$ the memory is smaller as $M_0-M_{\mathrm{BH}}\approx 1.05\mpl$, see Tab. \ref{table:params}.}
\label{fig:signals}
\end{center}
\end{figure}


\begin{figure*}[t]
\begin{center}
{\includegraphics[width=2\columnwidth]{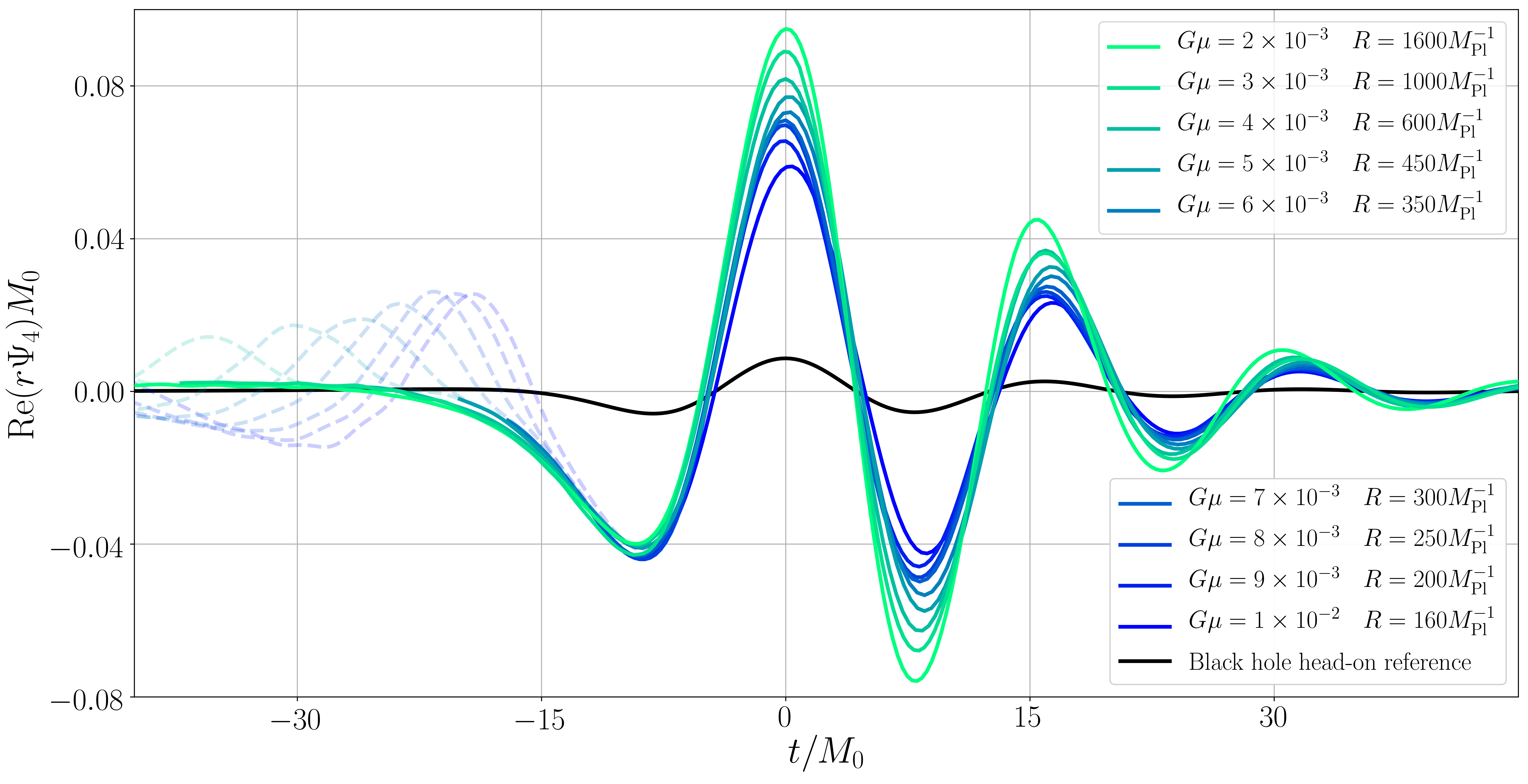}}
    \caption{{\bf Gravitational wave signal for different string tension $G\mu$} and black hole head-on reference \cite{Clough:2015sqa}: The signal is normalised with the initial mass of the system and shifted such that the maximum of $r\Psi_4$ coincides at time $t = 0$, for all cases in table (\ref{table:params}). Unphysical parts of the signal are de-emphasised using dashed lines. A summary of the parameters used for these runs is shown below in table (\ref{table:params}). } 
  \label{fig:allsignals}
  \end{center}
  \end{figure*}
  
  \begin{figure*}[t]\label{table:params}
\centering
\begin{tabular}{|c|c|c||c|c|c|c|c|}
\hline
$\qquad G\mu \qquad $            & $\quad R_0~[\mpl^{-1}] \quad$ &$\quad \lambda \quad $& $\quad M_0~[\mpl] \quad$ & $\quad M_\mathrm{BH}~[\mpl] \quad$ & $\quad E_\mathrm{matter}~[\mpl] \quad$ & $\quad E_\mathrm{GW}~[\mpl] \quad$ & $\qquad \gamma(t_\mathrm{BH}) \qquad$ \\ \hline \hline
$1\times 10^{-2}$ & $160$  & $~2~$          & $10.05\pm 0.07$                     &                      $9.21 \pm 0.18$ & $0.41\pm 0.10$              &$0.07 \pm 0.02$                              & $7.96$             \\ \hline
$9\times 10^{-3}$ & $200$  & $2$         & $11.31\pm 0.07$                    & $10.27 \pm 0.05$ & $0.31\pm 0.08$              & $0.09 \pm 0.02$                  & $8.84$            \\ \hline
$8\times 10^{-3}$ & $250$   & $2$         & $12.57\pm 0.07$                     &                      $11.59 \pm 0.08$ & $0.26\pm 0.07$              & $0.11 \pm 0.02$                  & $9.95$            \\ \hline
$7\times 10^{-3}$ & $300$   & $2$         & $13.19\pm 0.07$                    &                      $12.23 \pm 0.01$ & $0.29\pm 0.07$              &  $0.13 \pm 0.03$                  & $11.37$             \\ \hline
$6\times 10^{-3}$ & $350$   & $2$         & $13.19\pm 0.07$                     &                      $12.14 \pm 0.06$ & $0.46\pm 0.12$              &  $0.15 \pm 0.02$                  & $13.26$            \\ \hline
$5\times 10^{-3}$ & $450$   & $2$         & $14.14\pm 0.06$                    &                      $12.97 \pm 0.02$ & $0.56\pm 0.14$              &   $0.19 \pm 0.02$                  & $15.92$           \\ \hline
$4\times 10^{-3}$ & $600$   & $2$        & $15.08\pm 0.05$                    &                      $13.76 \pm 0.04$ & $0.75\pm 0.19$              & $0.25 \pm 0.02$                  & $19.89$            \\ \hline
$3\times 10^{-3}$ & $1000$  & $2$        & $18.85\pm 0.04$                     &                      $17.58 \pm 0.12$ & $0.62\pm 0.16$              &  $0.38 \pm 0.02$                  & $26.53$           \\ \hline
$2\times 10^{-3}$ & $1600$  & $2$        & $20.11\pm 0.03$                    &                      $18.33 \pm 0.06$ & $1.38\pm 0.35$               & $0.44 \pm 0.02$                  & $39.79$           \\ \hline
\end{tabular}
\caption{{\bf Overview of simulations with different $G\mu$ and $R_0$:} In this table, we list all the simulations we have done for this work.  The initial mass $M_0$ is obtained using \eqref{eq:mass} and the error calculated with the difference to the integrated mass of the numerical initial data. To extract the energy in gravitational waves $E_{GW}$ we integrated over the $r\Psi_4$ at different radii. The radiated energy in matter components $E_\mathrm{matter}$ is estimated by integrating it after black hole formation over the numerical grid excluding the interior of the BH.
 }
\end{figure*}


We first correct a time delay $\Delta t$ of the signal caused by
a redshift, which we estimate as
\begin{equation}\label{eq:delay}
\Delta t = \int_{R(t)}^{r_\mathrm{ext}}dr \left(\frac{1}{\alpha(t,r)}-1\right)
\end{equation}
where $\alpha(t,r)$ is the lapse function and
$\int_{R(t)}^{r_\mathrm{ext}}dr=r_\mathrm{ext} - R(t)$ is the distance from the
string center to the extraction radius as the loop collapses, which we track
throughout the simulation. The delay $\Delta t$ encodes the difference between
the simulation time and the real time it takes a gravitational wave to propagate
from the string center to the detector. The delay over time is shown in Fig.
\ref{fig:delay} and for near black hole formation, we estimate it to be $\Delta
t \approx 200\mpl^{-1}$ for $G\mu=4\times 10^{-3}$ and $R_0=600\mpl^{-1}$ case.

We then match the strain from our numerical relativity simulations $rh_{\text{num}}$ with the weak gravity calculation $rh_{\text{weak}}$ of the previous section as follows
\begin{equation}
rh = \begin{cases}
rh_{\text{weak}} , \quad & t < t_{cut}\\
rh_{\text{num}} + r\Delta h& t > t_{cut} \\
\end{cases}~.
\end{equation} 
The free shift $r\Delta h$ is chosen by finding the best fit value over a region where both signals are valid (shaded region in Fig. \ref{fig:fitting}). We define this region of validity as, that when $GM/R(t_f)\approx 0.25$, such that $t_f=R_0\cos^{-1}\left(4GM/R_0\right)$. In addition, we define the starting point as the time when most of the initial data artefacts have passed the detector (we can read this value from the $r\Psi_4$ plot). The best fit is shown in Fig. \ref{fig:fitting}.

A similar analysis can be done for other $(G\mu,R_0)$ cases, which we compare in Fig. \ref{fig:signals}. The larger the initial radius $R_0$ of the loop, the longer the infall. In addition, more energetic events (larger $E_\mathrm{GW}$) have larger amplitude $rh$ whereas we see that the last three waveforms show a similar amount of memory $r\Delta h$, which is expected as the total energy emitted is $M_0-M_{\mathrm{BH}} \approx 1.25 \mpl$, and for $(G\mu,R_0)=(6\times 10^{-3},350\mpl^{-1})$ less energy is radiated to infinity $M_0-M_{\mathrm{BH}} \approx 1.05 \mpl$, resulting in smaller memory, see Tab. \ref{table:params}.
 
\section{Summary of simulations}

Here we show the summary of the results for the different $G\mu$ runs. The first three columns correspond to the parameter space studied. The next four columns include information of data extracted from the simulations together with the corresponding error bars. Lastly, we compute the length contraction before black hole formation using the velocity given by the Nambu-Goto approximation (\eqref{eq:BHvel}).

To get an approximate estimate of the numerical precision of the signal in Figs. (\ref{fig:GmuRun}), (\ref{fig:DifferentLambda})  and (\ref{fig:DifferentRadii}), we performed two simulations with two different resolutions. Conservatively we can assume our simulations possess 2nd order convergence (see section \ref{sect:conv} below) and used the difference between the two runs to get an estimate for the error. We then chose the maximum value of the error (excluding the non-physical signal from the initial data) as the value for all points. 

Furthermore, in Fig. (\ref{fig:E_GW}) we calculated errors for all measured quantities. We estimated the error of $M_0$ by calculating the difference between the theoretical value and the integrated energy of the first frame. The errors for $M_{BH}$ are obtained by performing a best fit using QNMs after some different time.  To calculate $E_{\mathrm{matter}}$ we integrated over the grid, excluding a region close to the black hole.  Lastly, the error of $E_{\mathrm{GW}}$ is estimated by the energy in initial data artefacts mixed with the physical signal, ie. the energy between $t=R_0+r_\mathrm{ext}$ and when the artefacts have passed the detectors.

\section{Numerical Methodology}

The full numerical relativity initial data for the circular Abelian Higgs cosmic string loop is explained in our previous paper \cite{PhysRevD.99.104028}. We solve for $\chi$ using the Hamiltonian constraint. We reduce the spatial dimension of the problem by using its cylindrical symmetry. This solution is then further relaxed to obtain the final solution, which is that of an excited cosmic string loop. We then evolve using \textsc{GRChombo} \cite{Clough:2015sqa}, which solves the BSSN formulation of the Einstein equations \cite{PhysRevD.52.5428,PhysRevD.59.024007,PhysRevLett.96.111101}.

 \begin{figure}[t]
\begin{center}
{\vspace{0.25cm}
\includegraphics[width=1.0\columnwidth]{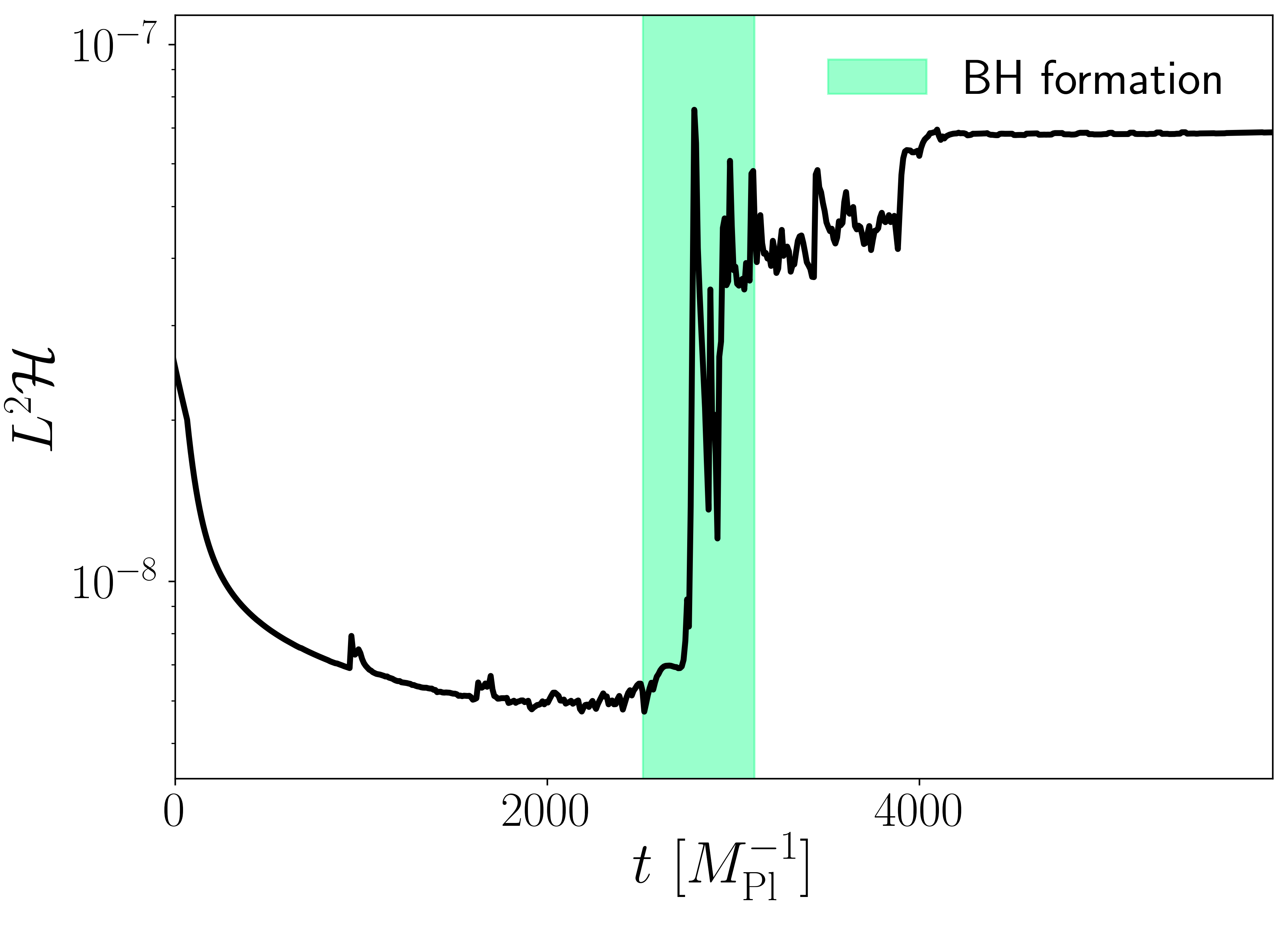}}
\caption{\textbf{$L^2$ norm of constraints:} We test the Hamiltonian constraint evolution for a loop with $G\mu=2\times 10^{-3}$ and $R_0=1600\mpl^{-1}$. It collapses and forms a black hole at $t\approx 2500\mpl^{-1}$. After that, the Hamiltonian constraint remains stable at $L^2\mathcal{H}<10^{-7}$. This plot shows that we have very good numerical control over our simulations.} 
\label{fig:L2H}
\end{center}
\end{figure}



\subsection{Numerical Extraction of Signal}

We extract the Penrose scalar $\Psi_4$ with tetrads proposed by \cite{Baker:2001sf}. Similarly as in black hole binaries, there is some non-physical radiation associated with the initial data, which in our case consists of a toroidal shell of artificial radiation resulting in two GW peaks before the physical signal. While such stray-GW can often be ignored as they quickly radiate away at light speed, due to the rapid collapse of the cosmic string loops at  ultrarelativistic speeds, they cannot be ignored.

\begin{figure}[t]
\begin{center}
{\includegraphics[width=1.0\columnwidth]{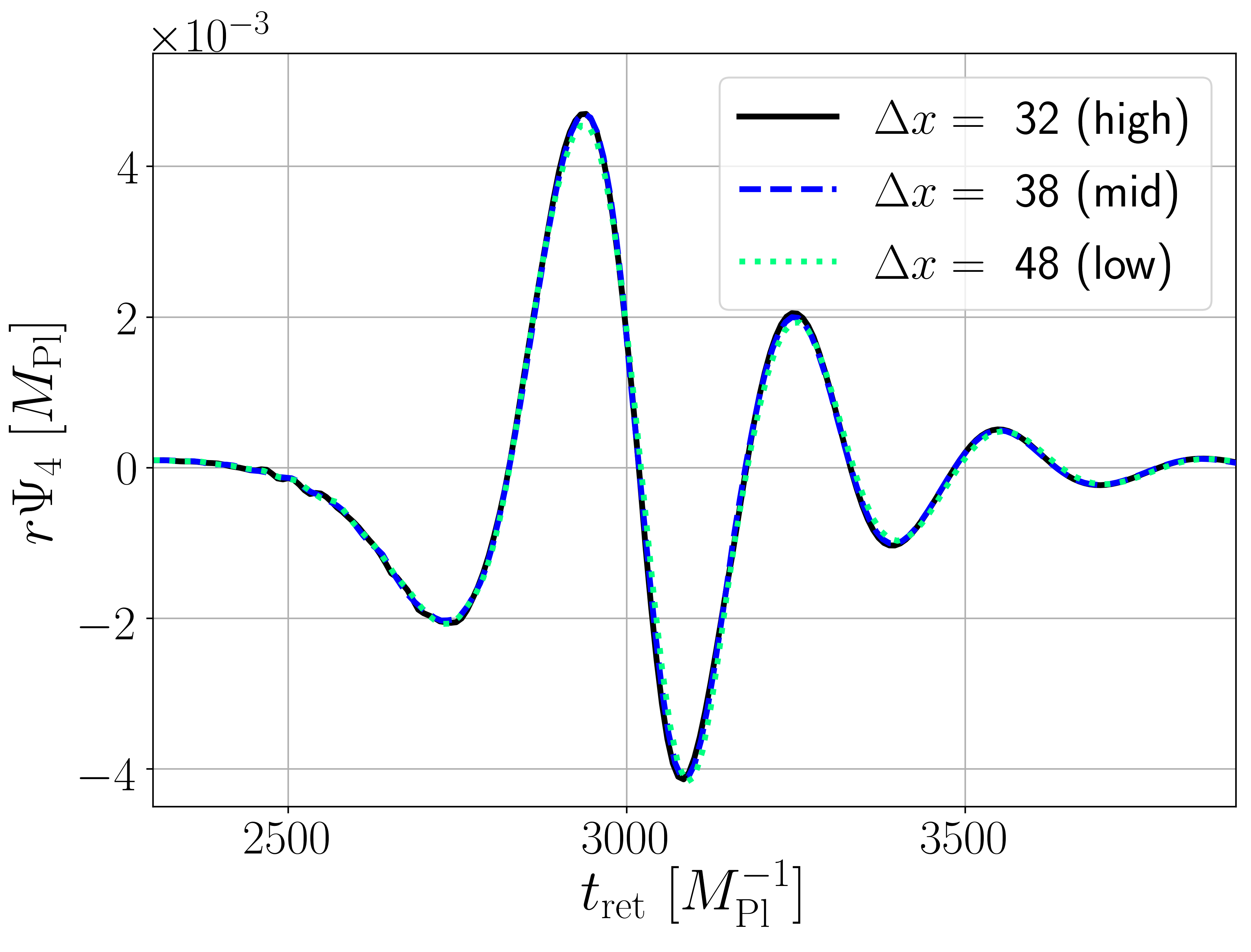}}
\vspace{-1.5em} \caption{\textbf{Convergence in $r\Psi_4$} for $G\mu=2\times 10^{-3}$ and $R_0=1600\mpl^{-1}$ between different coarse grid resolutions: low ($\Delta x=48\mpl^{-1}$), mid ($\Delta x=38\mpl^{-1}$) and high ($\Delta x=32\mpl^{-1}$) resolutions, in addition to $6$ refinement levels.}
\label{fig:Convergence}
\end{center}
\end{figure}

Nevertheless, these artefacts  can be separated by simulating larger loops. The time for the stray radiation moving at the speed of light is $R_0+r_\mathrm{ext}$, while the signal of the collapsing loop arrives around $R_0\pi/2+r_\mathrm{ext}$. This implies that we can separate the artificial radiation from the real signal by increasing the radius of the loop, which is computationally expensive. This is especially visible in Fig. \ref{fig:GmuRun}, where we increased the radius of the ring for smaller $G\mu$ to guarantee black hole formation. The initial peak, which is the artificial, becomes more and more separated with the signal for larger $R_0$. 

To calculate the total emitted GW energy we use the usual equation 
\begin{equation}\label{eq:energyPsi4}
\frac{dE_{\mathrm{GW}}}{dt} = \frac{r^2}{16\pi G}\int_{\mathcal{S}_r}\left|\int^t_{t_0}\Psi_4dt^{\prime}\right|^2 d\Omega~,
\end{equation}
where $\mathcal{S}_r$ is a sphere of radius $r$.

\subsection{Numerics and Convergence Tests} \label{sect:conv}

In Fig. (\ref{fig:L2H}), we show that the  volume-averaged Hamiltonian constraint violation
\begin{equation}\label{eq:L2H}
L^2(H) =\sqrt{\frac{1}{V} \int_V |\mathcal{H}^2| dV}~,
\end{equation}
where $V$ is the simulation box coordinate volume with the interior of the apparent horizon excised, is under control throughout the simulation.

We use the gradient conditions on $\phi$ and $\chi$ to tag cells for regridding. The precise criteria is chosen depending on the symmetry breaking scale $\eta$ and the total mass of the system. We use the symmetry of the system to only simulate one quarter of the system, which reduces the computational cost of the problem. 

We cut off our signal after some time $t$ when the black hole has formed (and hence the QNM signal is completely determined analytically), and fit QNM modes for the $l=2~m = 0$ mode \cite{Kokkotas:1999bd} in Fig. \ref{fig:QNMl2m0fit}). We test the precision of the simulation by comparing  the radiated energies with the initial mass. We find that these number for the simulations in table (\ref{table:params}) are consistent within the 1-5 \% range.

We tested the convergence of our simulations with a cosmic string loop of $G\mu = 2\times 10^{-3} $ and $R_0=1600\mpl^{-1}$ by using a box of size $L=3072\mpl^{-1}$ in which we improved by a factor of $1.2$ between the medium and highest resolution and $1.25$ between the lowest and medium resolution. The convergence of $r\Psi_4$ is shown in Fig. \ref{fig:Convergence}, for different coarse grid resolutions: low ($\Delta x=32\mpl^{-1}$), medium ($\Delta x=38.4\mpl^{-1}$) and high ($\Delta x=48\mpl^{-1}$), in addition to $6$ refinement levels.

\begin{figure*}[t]
\begin{center}
{\includegraphics[width=2\columnwidth]{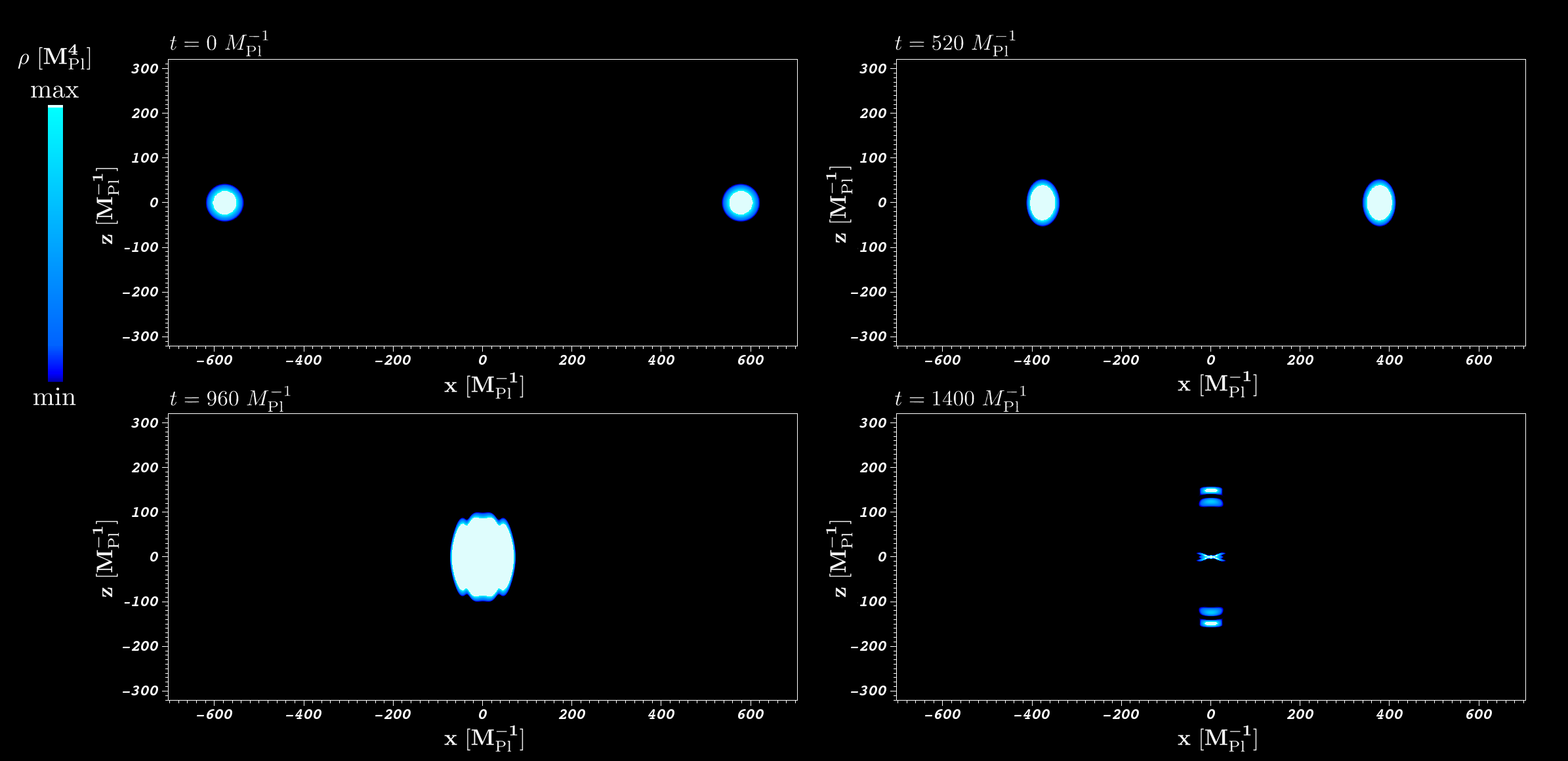}}
\caption{{\bf 2D slice of the collapse of a cosmic string loop using 3+1D numerical relativity.}: Figures in the panel above show the evolution of the system from left to right and top to bottom. In colour we plot the energy density.  Initially, the loop starts to collapse from rest (upper left); The energy density of the loop increases as its radius becomes shorter and accelerates to ultra-relativistic speeds, when Lorentz contraction effects emerge in the direction of the collapse (upper right). When the radius of the loop is of the same order as the width of the string, the collision happens, where high curvature effects appear (lower left). If the system is massive and thin enough, part of the initial mass of the system collapses to a black hole and high-relativistic jets are emitted axially as a result of the ultra-relativistic collision (lower right). This aspherical ejection of matter is responsible for a constant shift in the gravitational waveform known as gravitational wave memory. The full movie can be found  \href{https://youtu.be/0sSH54gXu4U}{here} \cite{Mattermovie}.}
\label{fig:panel}
\end{center}
\end{figure*}

\end{document}